\documentclass[aps,prx,preprint,superscriptaddress]{revtex4-2}

\usepackage{amsmath,amssymb}
\usepackage{bm,color}
\usepackage{graphicx}

\newcommand{\Fref}[1]{Fig.\ref{#1}}
\newcommand{\eref}[1]{eq.(\ref{#1})}

\begin{document}


\title{Evolutionary Shaping of Low-Dimensional Path Facilitates Robust and Plastic Switching Between Phenotypes}


\author{Ayaka Sakata}
\email[]{ayaka@ism.ac.jp}
\affiliation{Department of Statistical Inference \& Mathematics, The Institute of Statistical Mathematics, Tachikawa, Tokyo 190-8562, Japan}
\affiliation{Department of Statistical Science, The Graduate University for Advanced Science (SOKENDAI), Hayama-cho, Kanagawa 240-0193, Japan}
\author{Kunihiko Kaneko}
\affiliation{Center for Complex Systems Biology, Universal Biology Institute, University of Tokyo, Komaba, Tokyo 153-8902, Japan}
\affiliation{The Niels Bohr Institute, University of Copenhagen, Blegdamsvej 17, Copenhagen 2100-DK, Denmark}

\begin{abstract}
Biological systems must be robust for stable function against perturbations, 
but robustness alone is not sufficient. 
The ability to switch between appropriate states (phenotypes) 
in response to different conditions is essential for biological functions, 
as observed in allosteric enzymes and motor proteins. 
How are robustness and plasticity simultaneously acquired through evolution? 
In this study, we examine the evolution of 
genotypes that realize plastic switching between 
two endpoint phenotypes upon external inputs, 
as well as stationary expressions of phenotypes. 
Here, we introduce a statistical physics model consisting of spins, 
with active sites and regulatory sites, which are distinct from each other.
In our model, we represent the phenotype and genotype 
as spin configurations and the spin-spin interactions, respectively. 
The fitness for selection is given by the spin configuration,
whose behavior is governed by the genotypes.
Specifically,
the fitness for selection is given so that it takes a higher value 
as more of the active sites take two requested spin configurations 
depending on the states of the regulatory sites. 
The remaining spins do not directly affect the fitness, 
but they interact with other spins. 
We numerically evolve the matrices of spin-spin interactions (genotypes) by changing them with mutations and selection of those with higher fitness. 
Our numerical simulations show that characteristic genotypes with 
higher fitness evolve slightly above the phase transition temperature 
between replica symmetric and replica symmetry breaking phase 
in spin-glass theory. 
These genotypes shape 
two spin configurations separately depending on the regulation, 
where the two phenotypes are dominantly represented by
the genotypes' first and second eigenmodes, where
smooth switching of two phenotypes 
are achieved by following a one-dimensional, 
quarter-circle path connecting the two phenotypes. 
Upon changes in regulations, 
spin configurations are attracted to this path, 
which allows for robust and plastic switching between the two phenotypes. 
The statistical-physics analysis based on the two eigenmodes show that
the free energy landscape has a valley along the quarter-circle one-dimensional switching path. 
Robust attraction to the path is achieved through the evolution of 
interaction within non-active and non-regulatory spin sites, 
which themselves do not contribute to fitness. 
Our finding indicates that the compatibility of the robustness and plasticity is 
acquired by the evolution of the low-dimensionality in the phenotype space, 
which will be relevant to the understanding of the robust function of protein as well as the material design. 
\end{abstract}


\maketitle


\section{Introduction}
\label{sec:introduction}

Biological systems are inherently complex,
comprising numerous elements. 
Despite such complexity, 
they function robustly under environmental and stochastic perturbations. 
The biological function, in general, is given as a result of phenotypes, 
which are generated via dynamics based on genetic information. 
As a consequence, the function-related phenotypes need to be robustly shaped 
through the dynamics.
However, a single robust phenotype or fitted state is insufficient for a biological system to function under varying conditions. 
Phenotypes must exhibit plasticity, shifting to appropriate patterns in response to relevant signals or inputs
\cite{stout2005macrophages, pisarchik2014control}. 
For instance, the active sites of enzyme proteins can change 
between two conformations known as tense and relaxed states, 
induced by allosteric regulation 
\cite{changeux1961, MWC, changeux2013, Stefan2011}. 
Motor proteins, such as the myosin, kinesin, and dynein families, 
exhibit large-scale conformational changes in response 
to binding events \cite{Maragakis_Karplus2005, Karplus_Gao2004}. 
Phosphorylation of substrates in the mitogen-activated protein kinase cascades can  
switch between two states depending on modification by phosphatase or 
diphosphatase \cite{Markevich2004}. 
Gene expression pattern switches 
in response to signals are also necessary for cell survival. 
Thus, the ability to switch between appropriate phenotypes 
in response to different conditions is essential for biological functions.
Accordingly, the presence of multiple phenotypes and transitions among them 
in response to inputs must be shaped through evolution. 
Considering such changes in phenotypes, then,
plasticity to external conditions also needs to be 
required for the switching to different phenotypes, in addition to the robust expression of each phenotypes.
In general, 
how robustness and plasticity are compatible remains a fundamental question in 
biology \cite{Hatakeyama-Kaneko, Murugan2020}.

To study such plastic responses in biological systems, 
it is essential to understand the nature of switching pathways,
in addition to  
the multiple phenotypes corresponding to endpoint structures. 
An understanding of the switching pathways can 
aid in development of engineering techniques, such as drug design, that target the 
intermediate states of the switching pathways \cite{Laine2010,Laine2012}. 
Despite advances in structural biology in recent decades, 
molecular-level characterization of switching remains a challenge 
due to limitations in macromolecular X-ray crystallography, 
nuclear magnetic resonance, and small-angle X-ray scattering 
\cite{Panjkovich_Dmitri2016}. 
Hence, theoretical or numerical approaches are necessary 
to understand general characteristics of large-scale conformational switching \cite{Vaart2006, Orellana2019}. 
For instance, the plastic network model, an extension of the elastic network model \cite{Tirion1996,Bahar1997,Haliloglu1997}, 
was utilized to generate conformational switching pathways 
that are consistent with experimental data of the intermediate structures in Escherichia coli adenylate kinase 
\cite{Maragakis_Karplus2005}. 
The resulting pathways resemble combinations of low-energy normal modes obtained for the endpoint structures \cite{Maragakis_Karplus2005}. 
It has then been suggested that such preferred directionality may contribute to catalysis in many enzymes, achieving extraordinary rate acceleration and specificity \cite{Henzler_Wildman2007}. 
For Src kinase, such switching paths were explored 
by using a coarse-grained, two-state Go model, 
characterized by a two-dimensional free energy landscape \cite{Yand2008}. 

In general, theoretical and numerical methods to explore 
conformational changes 
assume the existence of probable switching paths, 
which minimize energy, free energy, or action \cite{Delarue2017}. 
The existence of a probable path implies that possible transient changes 
are constrained along the path. 
Further, low-dimensional approximations using principle component analysis 
have often been adopted to simplify the numerically or experimentally 
obtained switching paths \cite{Orellana2019}. 
These studies suggest the importance of understanding 
how low-dimensional switching paths are shaped and evolved
in the phenotypic spaces.

As for the stationary states,
recent experimental and numerical observations
have shown
that evolved phenotypes are often constrained within a low-dimensional manifold 
despite the high dimensionality of the phenotype space. 
For example, changes in (logarithmic) concentrations of mRNAs or proteins 
have been found to be correlated \cite{Barkai, MA1, Bahler,Matsumoto} 
or proportional \cite{KK-PRX, Heinemann} across all components 
under various environmental stresses. 
Numerical simulations of cell models with catalytic reaction networks 
have also demonstrated that evolved phenotypic changes caused 
by environmental and mutational changes are constrained 
within a low-dimensional manifold \cite{Furusawa-KK_PRE}. 
This reduction in dimensionality from high-dimensional phenotypes has also 
been observed in the structural changes of proteins, 
as a result of data analysis \cite{Tang}. 
Additionally, such dimensional reduction is suggested to be
a result of the robustness of phenotypes shaped by evolution.
However, such studies are
limited to phenotypes around the endpoint structures, i.e., the stationary conditions.
In this study, we examine the evolution of the switching path from the viewpoint of dimensional reduction. 

In particular, we address the following questions:
\begin{itemize}
\item Under what conditions and how are multiple endpoint phenotypes 
shaped depending on external inputs and stabilized through evolution?
\item Are low-dimensional constraints of switching paths shaped through evolution?
\item What are the characteristics of switching paths between endpoints?
\item What are the characteristics of evolved genotypes that allow robust switching paths?
\end{itemize}

To address these questions, we extend a spin-statistical physics model 
introduced previously \cite{Sakata-Kaneko2020}. 
In this model, the spin variables $\bm{S}$ and their interaction variables $\bm{J}$ 
represent phenotype and genotype \cite{SHK,SHK_PRE,SHK_EPL}, respectively,
and fitness is provided by certain spin configurations. 
We consider two endpoint structures, 
corresponding to those under regulation and without regulation. 
We introduce active and regulatory sites in the spin system 
to represent the effect of external regulation applied to the regulatory sites. 
The fitness of selective evolution depends on the appropriate 
expression of configurations. 
Fitted interactions can provide two configurations of active spins, 
corresponding to regulated and non-regulated cases. 

Numerical evolution allows us to 
examine how the robustness of each phenotype, 
as well as its plasticity to switch between the two configurations, 
is shaped by regulation.
Our result shows that, as a result of evolution, 
the dimensional reduction to a two-dimensional phenotype space appears, 
under a certain range of temperatures,
while a one-dimensional path is shaped for the switch 
between the two phenotypes in the regulated and non-regulated cases. 
The shaped switching path is robust to thermal noise and genetic mutation. 
In terms of statistical physics, the robustness of the fitted phenotype is achieved 
in the replica-symmetric phase. 
In contrast, the plasticity of the switch increases as the temperature approaches 
the replica-symmetry breaking (RSB) transition. 
We then will show that robust response is achieved near the RSB transition.

\section{Model}
\label{sec:model}
\begin{figure}
\centering
\includegraphics[width=6in]{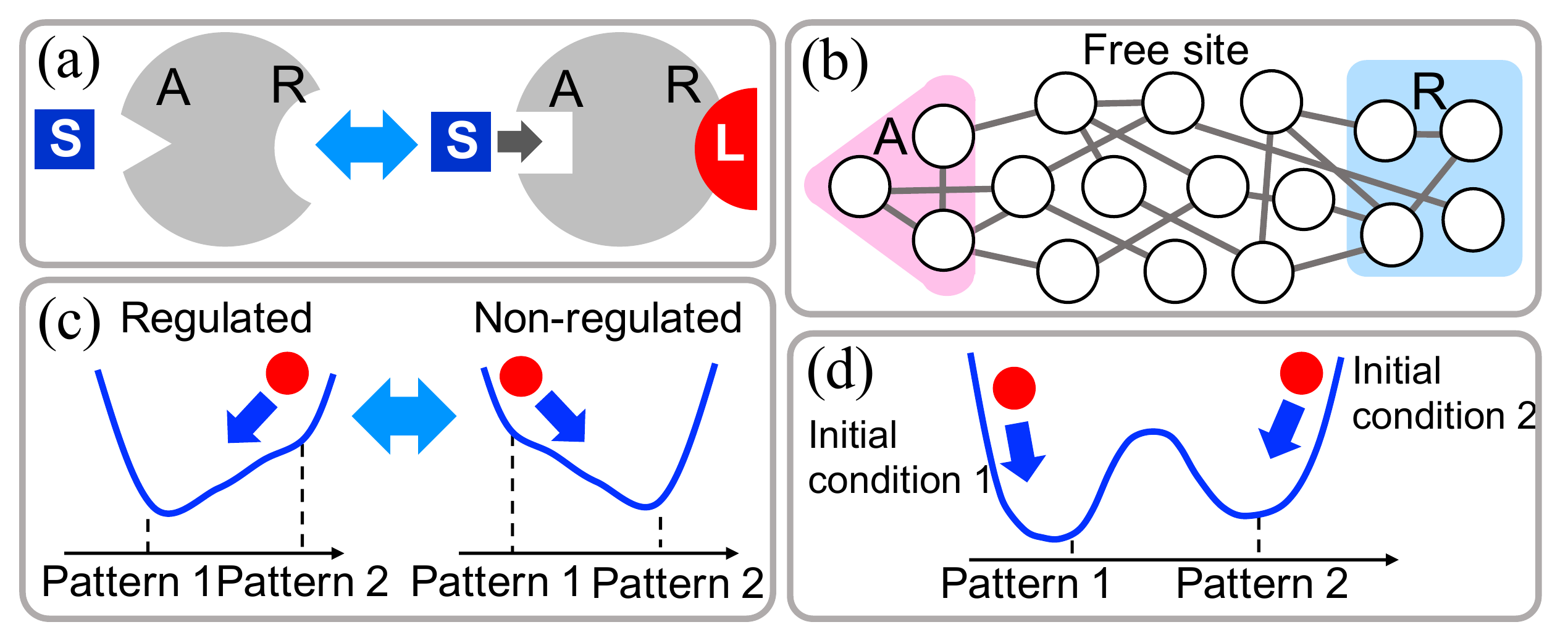}
\caption{(a) Schematic representation of the conformational change
induced by regulation.
A, R, and S denote active sites, regulatory sites, and substrates, respectively.
The molecule denoted by L is a ligand that regulates the protein
through regulatory sites. 
(b) The spin model for conformational switching
that has active and regulatory sites.
(c) A landscape picture of the 
conformational changes induced by regulation 
discussed in this study.
(d) Free energy landscape of the 
multi-pattern embedding in the associative memory model such as 
Hopfield networks. }
\label{fig:concepts}
\end{figure}
Here, we introduce an abstract model of
interacting spins, as a simplified representation
of proteins whose active sites conformation
is regulated by regulatory sites.
Fig. \ref{fig:concepts} (a) gives a 
simplified picture of the 
regulation and related conformational changes adopted in this study.
The protein shown as grey in the figure has an active (`A') and a
regulatory site (`R'), which are parts of the protein,
consisting of amino-acid residues.
In general allosteric regulation,
the active and regulatory sites
are located sufficiently apart and do not interact directly.
As shown in Fig. \ref{fig:concepts} (a),
binding of the ligand to the regulatory site leads to the
conformational change in the active site,
via interaction with sites other than the active sites and regulatory sites.
In contrast, without the binding of the activator to the regulatory site,
such conformational changes in the active site do not
occur and remain in its original conformation.

We introduce an abstract statistical-physics model with interacting spins
representing conformation, 
as shown in Fig.\ref{fig:concepts}(b).
The model consists of spin variable $\bm{S}=\{S_1,\cdots,S_N\}\in\{-1,+1\}^N$ which 
represents the conformational change in each amino-acid residue,
whereas the coupling $\bm{J}$ between spins represents the interaction among 
residues. 
These are respectively denoted by nodes and edges in Fig. \ref{fig:concepts}(b).
Here, we set $\bm{J}$ as $N\times N$ symmetric matrix.
Its elements are given by $J_{ii}=0~(i=1,\cdots,N)$,
and $J_{ij}\in\Omega_J$ for $i\neq j$ 
where $\Omega_J=\{-1\slash\sqrt{N},0,1\slash\sqrt{N}\}$.
Active (`A') and regulatory (`R') sites are represented by 
$N_A$ and $N_R$ spins among the $N$ spins.
The label sets of active and regulatory spins are 
denoted by ${\cal A}$ and ${\cal R}$, respectively. 
We set $J_{ij}=0$ for $i\in{\cal A}$ and $j\in{\cal R}$
or $i\in{\cal R}$ and $j\in{\cal A}$, to 
prohibit the direct interaction between regulatory and active sites.
The spin variables other than those at the active and regulatory sites 
are called free sites, as shown in Fig. \ref{fig:concepts} (b).

For the dynamics of spin-variables under given $\bm{J}$, we adopt 
the transition rule of spins from $\bm{S}$ to $\bm{S}^\prime$
under given 
Hamiltonian with the interaction matrix $\bm{J}$
and temperature $T$ as
\begin{eqnarray}
\mathrm{Pr}[\bm{S}\to\bm{S}^\prime|\bm{J}]=\min\{\exp(-\beta\Delta_H(\bm{S},\bm{S}^\prime|\bm{J})),1\},
\label{eq:S_expression}
\end{eqnarray}
where $\beta=T^{-1}$ is the inverse of temperature, and
$\Delta_H(\bm{S},\bm{S}^\prime|\bm{J})\equiv H(\bm{S}^\prime|\bm{J})-H(\bm{S}|\bm{J})$.
Here, we set the Hamiltonian as 
\begin{eqnarray}
H(\bm{S}|\bm{J})=-\frac{1}{2}\bm{S}^{\top}\bm{J}\bm{S},
\label{eq:Hamiltonian}
\end{eqnarray}
where $\top$ denotes the matrix transpose.

Note that in this statistical-physics model, 
we adopt the spin variables $\{-1,1\}$, 
instead of continuous conformational variables in residues. 
This is a highly simplified model by nature 
(see \cite{Saito-Sasai-Yomo}
for examples of spin models for protein dynamics). 
Here, we aim to elucidate how certain stochastic dynamics for generating functional 
phenotypes are shaped through evolution. To this end, the present model capture the 
essence of such dynamics and genotype-phenotype mapping, in which spin variables 
$\bm{S}$ 
corresponding to the phenotypes are shaped by high-dimensional dynamics under 
genetic rules given by the interaction matrix $\bm{J}$, whereas regulation is 
referred to as 
change in a part of ``regulatory"
spins, as defined below.


Next, the functional change in the active sites is postulated by the appropriate change in the configuration of active spins $\bm{S}_{\cal A}=\{S_i|i\in{\cal A}\}$, depending on 
the configurations of the regulatory site $\bm{S}_{\cal R}=\{S_i|i\in{\cal R}\}$.
Here, instead of introducing the binding of ligands to regulatory sites 
as external variables, we assume that the configuration of the spins is set at 
$\bm{S}_{\cal R}^+$ upon the binding. That is, among $2^{N_R}$ possible configurations of the regulatory spins, the regulatory spins only take the configuration
in $\bm{S}_{\cal R}^+$ when the ligand binding occurs. 
Further, we consider that 
$\bm{S}_{\cal R}^+$ 
cannot appear without the ligand binding.
Accordingly, the equilibrium distribution upon the regulation and non-regulation is given by
\begin{align}
P_\beta^+(\bm{S}|\bm{J})=\frac{1}{Z_\beta^+}\exp(-\beta H(\bm{S}|\bm{J})),~~Z_\beta^+(\bm{J})=\sum_{\bm{S}|\bm{S}_{\cal R}\in\bm{S}_{\cal R}^+}\exp(-\beta H),\label{eq:dist_pos}\\
P_\beta^-(\bm{S}|\bm{J})=\frac{1}{Z_\beta^-}\exp(-\beta H(\bm{S}|\bm{J})),~~~Z_\beta^-(\bm{J})=\sum_{\bm{S}|\bm{S}_{\cal R}\notin\bm{S}_{\cal R}^+}\exp(-\beta H),
\label{eq:dist_neg}
\end{align}
where $\bm{S}|\bm{S}_{\cal R}\in\bm{S}_{\cal R}^+$ 
and $\bm{S}|\bm{S}_{\cal R}\notin\bm{S}_{\cal R}^+$
indicate the set of possible configurations 
for regulated and non-regulated states, respectively.


Next, the functional change in configurations of the active spins in response to the 
regulation is given by the change in regulatory spins from 
$\bm{S}_{\cal A}^-$ to $\bm{S}_{\cal A}^+$: 
Thus, the conformational change induced by regulation is modeled as follows:
if the configuration of regulatory spins is set at $\bm{S}_{\cal R}^+$,
the configuration of the active spins turns into $\bm{S}_{\cal A}^+$;
otherwise, the configuration of the active sites stays at $\bm{S}_{\cal A}^-$.

The function of the present system
to express the target spin pattern $\bm{S}_{\cal A}^\pm$ appropriately
can be measured by the 
magnetization $m_A^\pm$
defined as the overlap of the spins in the active sites 
with the corresponding target spin patterns as 
\begin{align}
m_A^+&=\frac{1}{N_A}\sum_{i\in{\cal A}}S_iS_i^+\\
m_A^-&=\frac{1}{N_A}\sum_{i\in{\cal A}}S_iS_i^-.
\end{align}
Finally, the overall
fitness that measures the functionality of the present system
is given by the sum of the expectations of $m_A^\pm$ as 
\begin{align}
\psi(\bm{J})=\frac{1}{2}\left\{\langle|m_A^+|\rangle_++\langle|m_A^-|\rangle_-\right\},
\label{eq:fitness_def}
\end{align}
where $\langle\cdot\rangle_+$ and $\langle\cdot\rangle_-$
are the expectation values according to the equilibrium distributions for regulated and non-regulated cases \eref{eq:dist_pos} and \eref{eq:dist_neg},

The evolution of genotypes $\bm{J}$ is then based on the above fitness $\psi(\bm{J})$. 
Higher fitness genotypes are selected under given selective pressure: At generation $g$, 
the evolutionary change in $\bm{J}$ to increase the fitness is given by
\begin{eqnarray}
\mathrm{Pr}[\bm{J}^{(g)}\to\bm{J}^{(g+1)}]=\min\{\exp(\beta_J\Delta\psi),1\},
\end{eqnarray}
where $\Delta\psi=\psi(\bm{J}^{(g+1)})-\psi(\bm{J}^{(g)})$.
The parameter $\beta_J=T_J^{-1}$ expresses the selection pressure, and the 
genotypes are selected uniformly at high temperatures $T_J\to\infty$, 
whereas at low $T_J$, 
genotypes with higher fitness values are preferred.

Remark: The celebrated Hopfield neural network model can be used for embedding several patterns in spin models.
In this case, as schematically shown in Fig. \ref{fig:concepts} (d),
multiple patterns with different spin configurations were reached depending on the initial 
condition given by Hamiltonian dynamics. In contrast, in our case, by external inputs to 
regulatory spins (i.e., with inputs or with different boundary conditions), 
different spin configurations are reached depending on if the regulatory spins are regulated or not, from the same initial conditions for the two, 
as in Fig.\ref{fig:concepts} (c), which has been introduced in the 
study of the reshaping of the energy landscape of a protein by allostery \cite{Swain2006}.
(In the context of the neural network model, this corresponds to the associative memory model upon external inputs \cite{Kurikawa-Kaneko2021}.)

\section{Numerical simulation}

Without loss of generality,
we set the indices of the regulatory sites and active sites as 
${\cal R}=\{N-N_R+1,\ldots,N\}$ and 
${\cal A}=\{1,\ldots,N_{\mathrm{A}}\}$, respecively.
Further, we set the configuration $\bm{S}_{\cal R}^+$
as $\bm{S}_{\cal R}^+=\{\{+1,\cdots,+1\},\{-1,\cdots,-1\}\}$.
For the desirable configurations of the active sites,
we set $\bm{S}_A^+=\{\{+1,\cdots,+1\},\{-1,\cdots,-1\}\}$
and $\bm{S}_A^-=\{\{+1,-1,\cdots\},\{-1,+1,\cdots\}\}$.
In the genotype evolution process,
we induce a 10-point mutation at each generation 
to generate the candidate of the next generation $\bm{J}^{\prime}$ from $\bm{J}$,
maintaining the symmetry ${\bm{J}^{\prime}}^\top=\bm{J}^{\prime}$.
Here, we mainly show the results for $N=100$,
$N_A=5$ and $N_R=10$, and the free sites consist of 
$N-N_A-N_R=85$ spin variables.
We update $\bm{J}$ 
at a sufficiently large value as $\beta_J=100$,
and discuss $T$-dependencies.

\subsection{Fitness, rugged landscape, and separation of two patterns}

\begin{figure}
\centering
\includegraphics[width=5.5in]{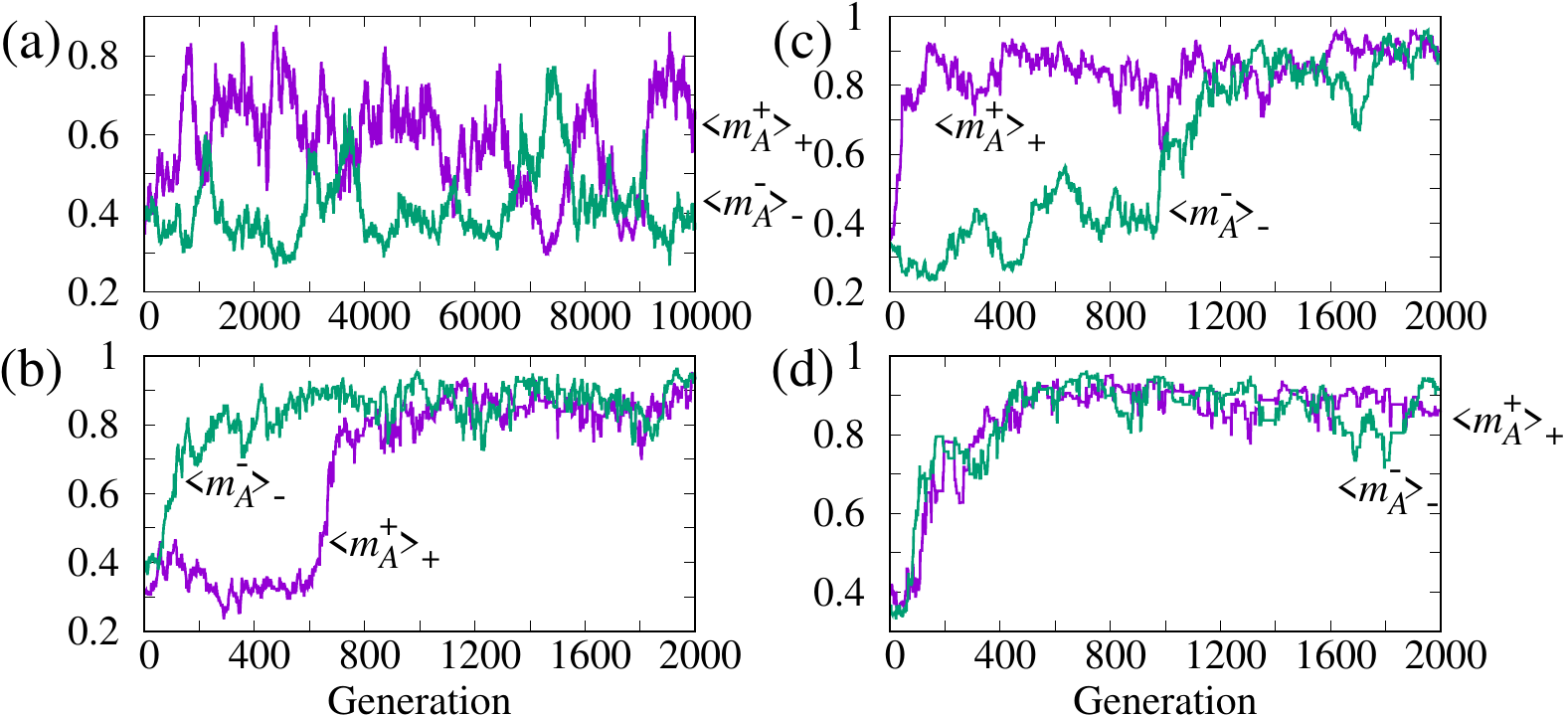}
\caption{Evolutional dynamics of $\langle m_A^\pm\rangle_\pm$
associated with the evolution of a genotype $\bm{J}$.
An example at $T=0.91$ is shown in (a),
and examples at $T=0.67$ are shown in 
(b)-(d).}
\label{fig:m_vs_gen}
\end{figure}

In Fig.\ref{fig:m_vs_gen},
we show examples of evolutionary dynamics
of $\bm{J}$ through the evolutionary changes of 
$\langle |m_A^\pm|\rangle_\pm^{(g)}$,
where $\langle\cdot\rangle^{(g)}$ denotes the expectation 
according to the distribution 
$P_\beta^\pm(\bm{S}|\bm{J}^{(g)})$
with a $g$-th generation genotype
$\bm{J}^{(g)}$. 
These quantities measure the tendency to exhibit 
desirable patterns depending on the regulatory site,
whereas fitness is given by their mean, as eq.(\ref{eq:fitness_def}).
Fig.\ref{fig:m_vs_gen} (a) shows an example of 
the generational changes of $\langle |m_A^\pm|\rangle_\pm^{(g)}$
at $T=0.91$.
They show a negative correlation; when $\langle|m_A^+|\rangle_+^{(g)}$ increases, 
$\langle |m_A^-|\rangle_-^{(g)}$ decreases,
and vice versa.
For the genotypes that show this behavior,
which are the most evolved genotypes 
around $T=0.91$, 
the simultaneous expression of $\bm{S}_A^+$
and $\bm{S}_A^-$, depending on the regulatory sites,
is difficult.
When the active sites take
one of the configurations in $\bm{S}_A^+$ or $\bm{S}_A^-$,
irrespective of the regulatory sites, 
we obtain $|m_A^+|=1$ and $|m_A^-|=0.2$
or $|m_A^+|=0.2$ and $|m_A^-|=1$, respectively.
Therefore, the fitness value of the genotype
that can express only one desirable pattern among $\bm{S}_A^\pm$
can reach a highest value of 0.6.
Meanwhile, 
at a lower temperature $T=0.67$, both
$\langle |m_A^\pm|\rangle_\pm^{(g)}$
increase simultaneously after the evolution,
as shown in Fig.\ref{fig:m_vs_gen}(b)-(d),
where the fitness reach around 0.9.
There are three evolutionary courses;
$\langle|m_A^+|\rangle_+^{(g)}$ or $\langle|m_A^-|\rangle_-^{(g)}$
increases first (Fig.\ref{fig:m_vs_gen}(b) or (c)),
or they increase simultaneously (Fig.\ref{fig:m_vs_gen}(d)).
Among 100 samples,
21, 33,
and 46 samples follow each course, respectively.

\begin{figure}
\centering
\includegraphics[width=2.7in]{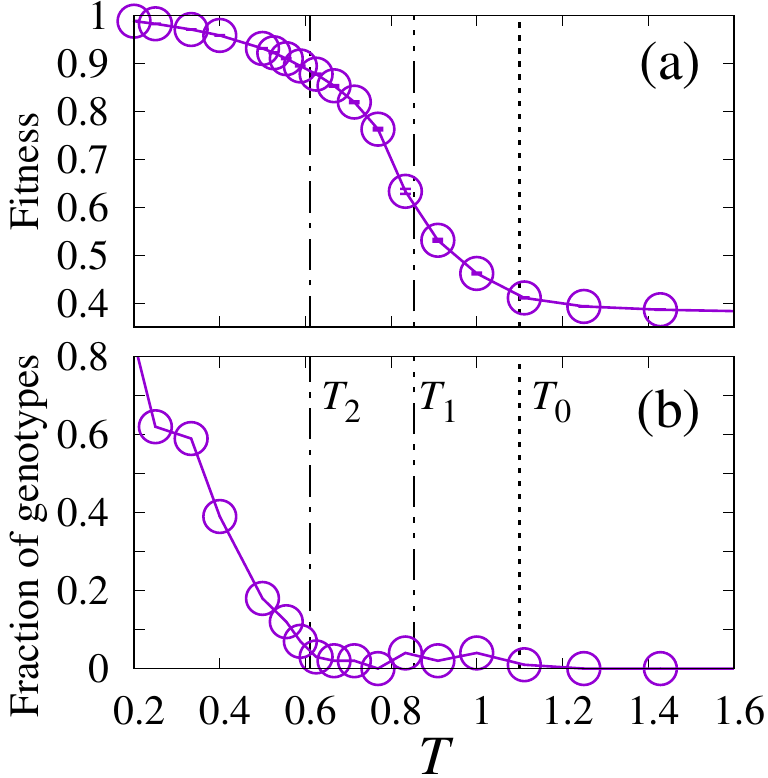}
\includegraphics[width=2.75in]{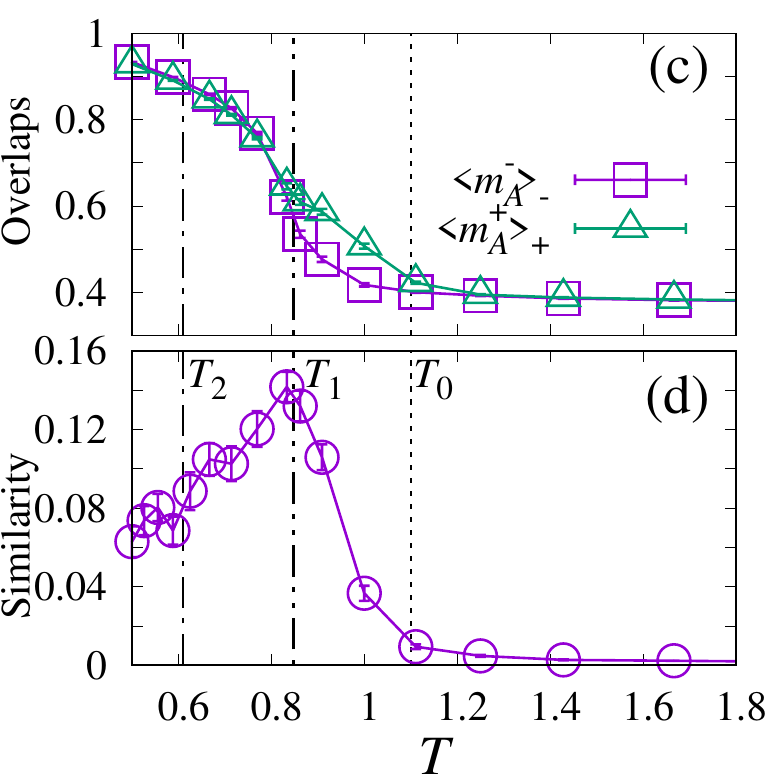}
\caption{$T$-dependence of (a) fitness,
(b) the fraction of genotypes on which the BP algorithm does not converge,
(c) $\langle m_A^+\rangle_+$ and $\langle m_A^-\rangle_-$,
and (d) Similarity between regulated and non-regulated states.
Each data point is averaged over 100 samples of evolved $\bm{J}$.
The vertical dotted line,
two-dot chain line, and one-dot chain line
denote $T_{0}$, $T_{1}$, and $T_{2}$, respectively.}
\label{fig:fitness}
\end{figure}

For each $T$, we obtain 100 samples of the
evolved $\bm{J}$ updated for 
$g=10^5$ generations,
and denote the temperature-dependence 
ensemble of evolved genotypes as ${\cal J}(T)$.
\Fref{fig:fitness} (a) shows the $T$-dependence of the 
mean of the fitness among ${\cal J}(T)$.
With a decrease in $T$, the fitness value $\Psi$ 
increases from 0.4,
which is a trivial value given by the uniform distribution of the phenotype $\bm{S}$.
$T_0$ is defined as the transition temperature 
characterized by fitness
below, in which the fitness value increases as $T$ decreases.
To be precise, it is defined by the point 
where the second derivative of $\Psi$ shows discontinuity.
We term the phase $T>T_0$ as the paramagnetic phase.
Next,
the energy landscape on $\bm{J}\in{\cal J}(T)$
governing the phenotype expression dynamics
changes at $T=T_2$:
We term the phases at $T_0>T>T_2$ and $T<T_2$ as replica-symmetric (RS) phase
and replica symmetry breaking (RSB) phase, respectively.
The difference between the two phases can be detected by 
the belief propagation (BP) algorithm \cite{Pearl,PGM,MM}.
In the fully connected spin-glass system,
the stability condition of the
BP algorithm agrees with the validity of the RS assumption in the replica analysis,
which is known as de Almeida-Thouless (AT) instability \cite{AT,Kabashima_CDMA};
hence, when the BP algorithm converges, the system on $\bm{J}$ 
corresponds to the RS phase, otherwise the RSB phase.
The RSB indicates the rugged landscape
with exponential orders of metastable states, and 
the phenotype expression dynamics is not robust to 
thermal fluctuation \cite{SHK}\footnote[1]{
While the AT instability cannot be derived from not fully connected systems,
including our evolutionary model, the BP algorithm is often used as a
numerical method to identify the phase of not fully connected systems.}.
At $T>T_2$,
most of the evolved genotypes in ${\cal J}(T)$ exhibit 
rapid convergence of the BP algorithm.
Meanwhile,
the BP algorithm cannot converge 
for most evolved genotypes in ${\cal J}(T)$ when $T<T_{2}$.
In \Fref{fig:fitness}(b),
we present the fraction of evolved genotypes for which
the BP algorithm does not appear to converge within $10^5$ steps,
which increases as $T$ is decreased below $T_2$.

%

The existence of these transitions from the paramagnetic phase to the RS phase, 
and then to the RSB phase,
is common with
the evolving spin-glass model to express one specific phenotype 
\cite{SHK,SHK_PRE,SHK_EPL,Sakata-Kaneko2020}.
In the present model,
however, another transition appeared at $T_1$,
with respect to the achievability of 
two patterns.
In Fig.\ref{fig:fitness}(c),
we show the temperature dependence of the overlaps
$\langle |m_A^+|\rangle_+$ and $\langle |m_A^-|\rangle_-$,
whose mean corresponds to fitness.
At $T_1<T<T_0$,
$\langle |m_A^+|\rangle_+$ contributes more to fitness,
and only the phenotype expression with regulation is preferentially shaped.
At $T< T_1$, both the increase of 
$\langle |m_A^-|\rangle_+$ and $\langle |m_A^-|\rangle_-$ 
are achieved depending on the regulatory site.
We term the phases $T_0>T>T_1$ and $T_1>T>T_2$
as RS1 and RS2, respectively.
Here, we note a 
negative correlation between $\langle m_A^\pm\rangle_\pm^{(g)}$
is observed 
in the paramagnetic and RS1 phases,
as shown in Fig.\ref{fig:m_vs_gen}(a).
In the RS2 and RSB phases,
the increase of both $\langle m_A^\pm\rangle_\pm^{(g)}$
is achieved
after a sufficient update,
as shown in Fig.\ref{fig:m_vs_gen}(b)-(d).

To study the transition at $T=T_1$,
we examine the 
probability distributions of spin configurations,
$p_\beta^+(\bm{S}|\bm{J})$ and $p_\beta^-(\bm{S}|\bm{J})$,
with and without regulations,
by means of
the component-wise expected phenotype 
for each $i=1,\ldots,N$ defined as
\begin{align}
\mu_i^\pm=\left\langle\mathrm{sign}\left(\sum_{i=1}^{N_A}S_i\right)S_i\right\rangle_\pm,
\end{align}
where the term $\mathrm{sign}(\sum_{i=1}^{N_A}S_i)$
is introduced to break the Z2 symmetry.
In Fig.\ref{fig:fitness}(d),
we show $T$-dependence of 
the similarities between two mean phenotypes
measured by
$\sum_{i=1}^N\mu_i^+\mu_i^-\slash N$.
As shown in Fig.\ref{fig:fitness}(d),
the overlap shows a peak at $T=T_1$,
and it decreases as $T$ decreases below $T_1$.
According to the decrease in the overlap, the transition 
between the phenotype with and without regulation
involves large conformational changes.
%

\subsection{Two-dimensional structure in the phenotype space}

\begin{figure}
\begin{minipage}{0.495\hsize}
\centering
\includegraphics[width=4in]{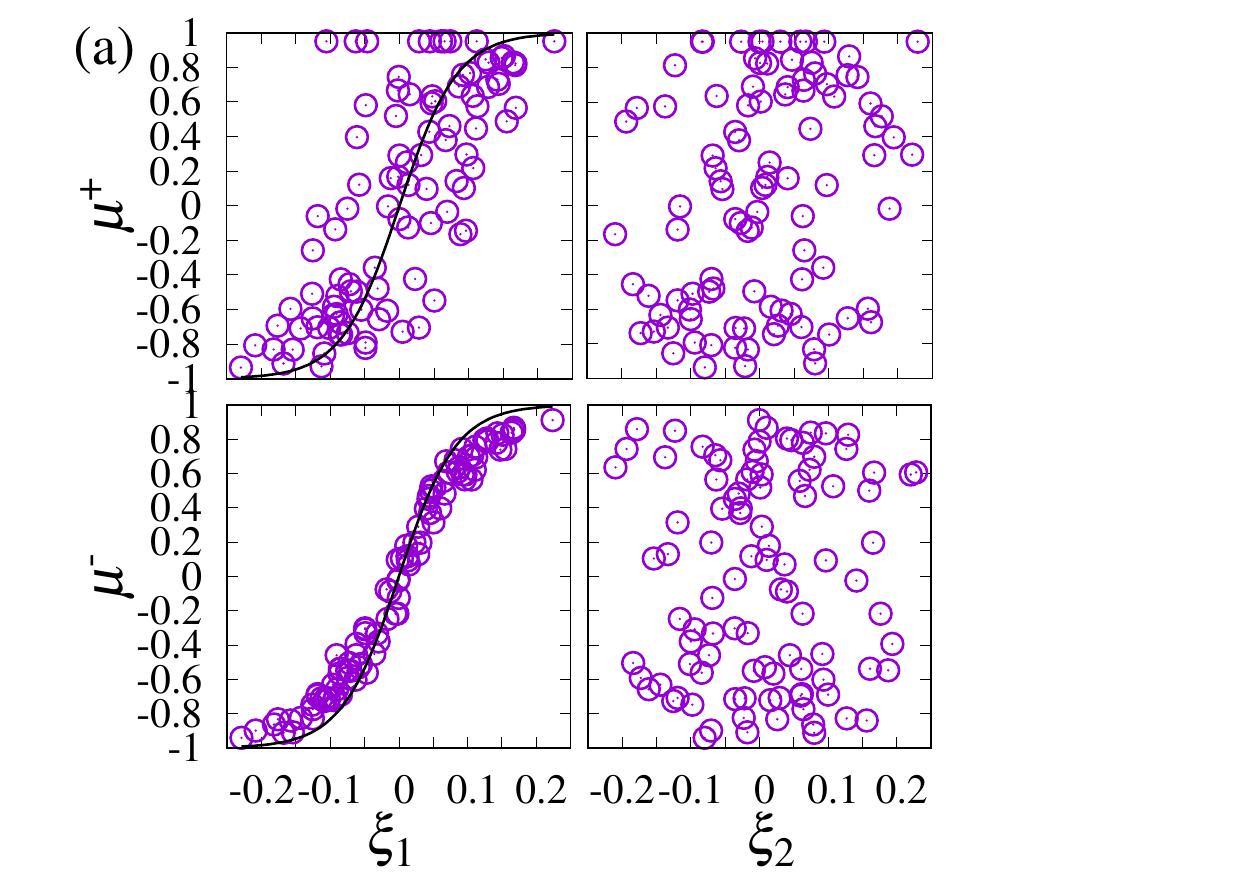}
\end{minipage}
\begin{minipage}{0.495\hsize}
\centering
\includegraphics[width=4in]{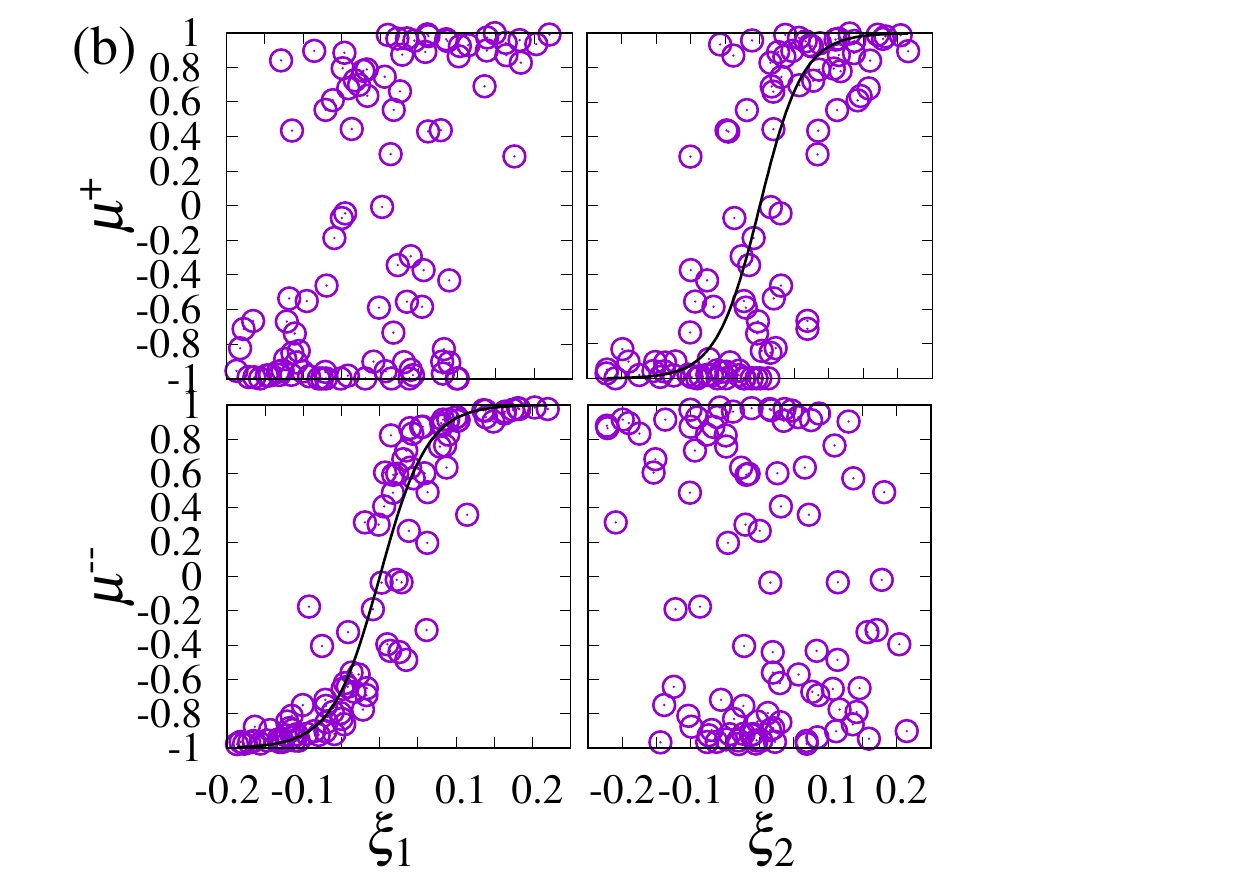}
\end{minipage}
\caption{Comparison between the eigenvectors ($\bm{\xi}_1$ and $\bm{\xi}_2$)
and $\bm{\mu}^\pm$ for 
one sample of evolved $\bm{J}\in{\cal J}(T)$ 
at (a) $T=0.833$ and (b) $T=0.667$.
The solid lines show the relationship $y=\tanh(\beta\sqrt{N}x)$.}
\label{fig:mu_vs_xi}
\end{figure}

We investigate how the two patterns shaped by evolution
are separated in the RS2 phase.
To compare $N$-dimensional 
mean phenotypes $\bm{\mu}^+$ under 
and $\bm{\mu}^-$ without regulation,
it is convenient to determine a reference coordinate system.
We adopte the eigenvectors of the evolved genotypes
as the axes to represent mean phenotypes.
Using the eigenvectors and corresponding eigenvalues,
the genotype $\bm{J}$ is decomposed as 
\begin{align}
\bm{J}=\sum_{r=1}^N\lambda_r\bm{\xi}_r\bm{\xi}_r^{\top},
\end{align}
where $\bm{\xi}_i$ and $\lambda_i$ are $i$-th eigenvector
and $i$-th eivenvalue.
We set the indices of the eigenmodes to be 
$\lambda_1\geq\lambda_2\geq\cdots\geq\lambda_N$.

In Fig. \ref{fig:mu_vs_xi},
we show the scatter plots between $\mu_i^\pm$ against eigenvectors $\xi_{1i}$ and 
$\xi_{2i}$ for $i=1,\ldots,N$
at (a) $T=0.833$ (RS1 phase) and 
(b) $T=0.667$ (RS2 phase)
under one realization of $\bm{J}\in{\cal J}(T)$.
In the RS1 phase,
the mean phenotypes $\bm{\mu}^\pm$,
in particular $\bm{\mu}^-$, are highly correlated with
$\bm{\xi}_1$,
as described by $y=\tanh(\beta\sqrt{N}x)$ (see Fig. \ref{fig:mu_vs_xi}(a)).
Here, the function $\tanh$ is consistent with the 
mean-field form of the magnetization
$\mu_i^\pm=\tanh(\beta \sum_{j\neq i}J_{ij}\mu^\pm_j)$.
Meanwhile, in the RS2 phase,
the regulated $\bm{\mu}^+$
and non-regulated $\bm{\mu}^-$
states exhibit correlations with $\bm{\xi}_1$ and $\bm{\xi}_2$,
respectively,
as shown in Fig.\ref{fig:mu_vs_xi}(b).
In both phases, the correlations between $\bm{\mu}^\pm$
and $\bm{\xi}_r$ ($r\geq 3$) are negligible.

\begin{figure}
\centering
\includegraphics[width=2in]{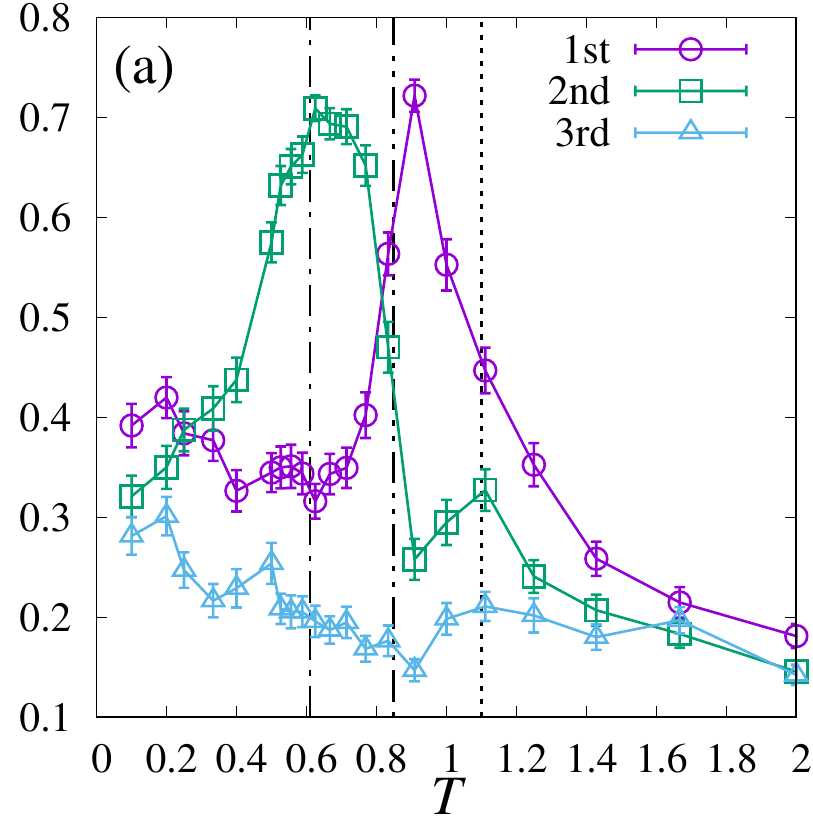}
\includegraphics[width=2in]{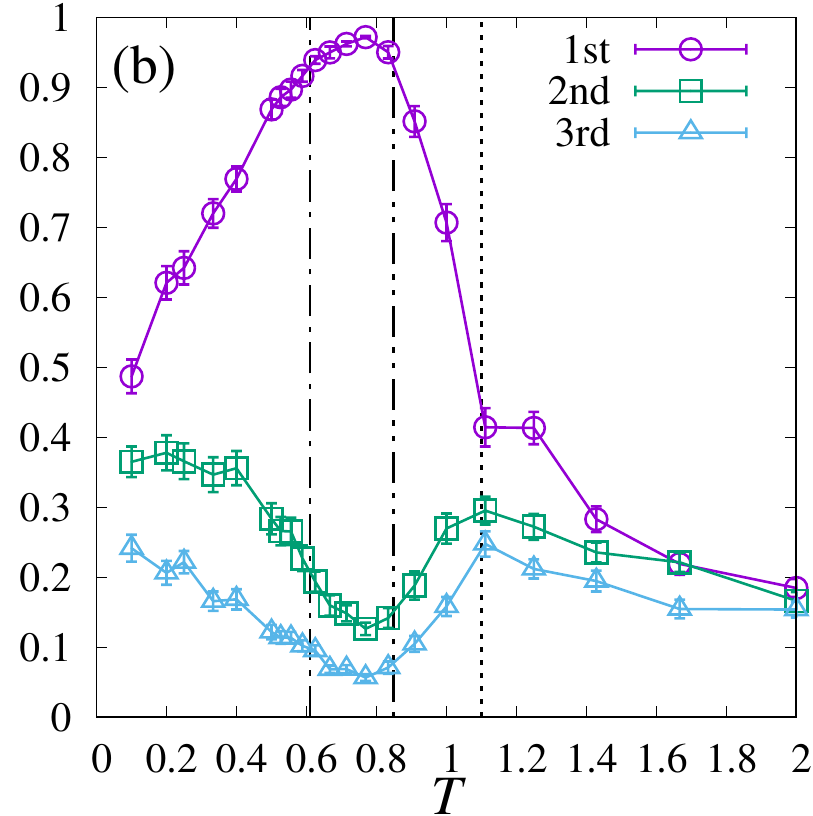}
\includegraphics[width=2in]{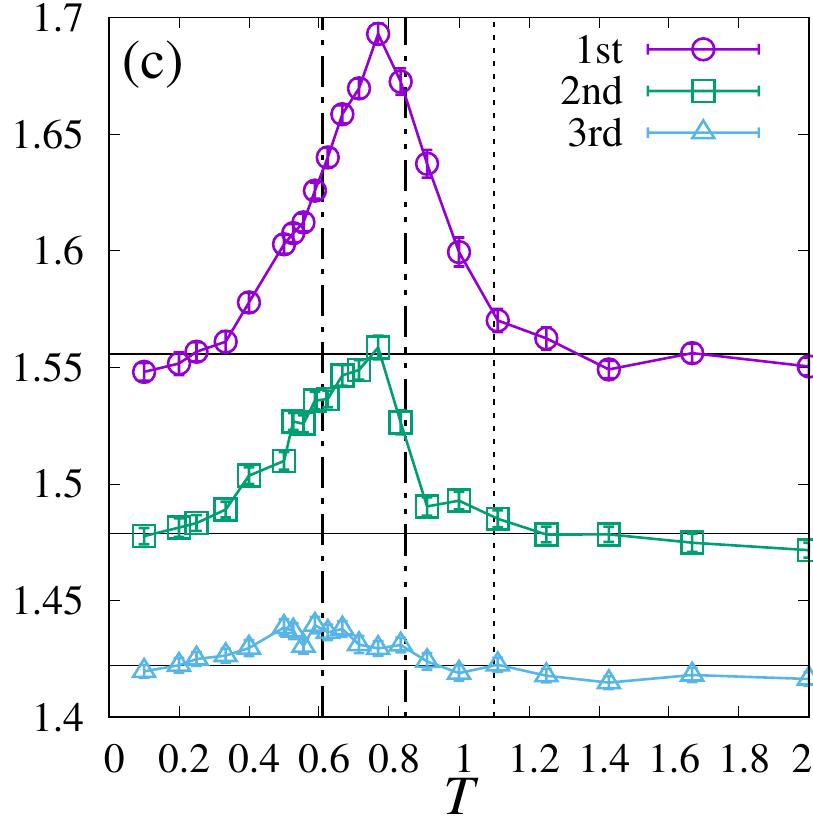}
\caption{The correlation coefficient between the first, second, and third eigenvectors with 
(a) regulated state $\bm{\mu}^+$ and (b) non-regulated state $\bm{\mu}^-$.
The $T$-dependence of the first, second, and third eigenvalues 
of the evolved genotypes is shown in (c),
where the three horizontal lines represent the expected eigenvalues
for randomly generated $\bm{J}$.
The vertical dashed line, one-dot-chain line, and two-dot-chain line
denote $T_{0}$, $T_{2}$ and $T_{1}$,
respectively.
Each point is averaged over 100 samples of the evolved $\bm{J}$.}
\label{fig:corr_eig_and_m}
\end{figure}

In \Fref{fig:corr_eig_and_m} (a) and (b),
we show the temperature dependence of the 
correlation between the eigenvectors and $\bm{\mu}^\pm$
by introducing the correlation coefficient 
between $\{\xi_{ri}\}$ and $\{\mathrm{atanh}(\mu_i^\pm)\}$
for $r=1,2,3$.
Here, the function $\mathrm{atanh}$ is introduced 
by considering the 
$\tanh$-form dependencies of 
$\bm{\mu}^\pm$ on $\bm{\xi}_1$ or $\bm{\xi}_2$, as shown in 
Fig.\ref{fig:mu_vs_xi}.
We denote the vector consisting of $\mathrm{atanh}(\mu_i^\pm)~(i=1,\cdots,N)$
as $\mathrm{atanh}(\bm{\mu}^\pm)\in\mathbb{R}^N$.
As shown in \Fref{fig:corr_eig_and_m} (a),
the correlation coefficient between 
the first eigenvector $\bm{\xi}_1$
and the mean phenotype with regulation $\bm{\mu}^+$
increases at $T<T_{0}$, namely in the RS1 phase.
As the temperature is lowered further below $T_{1}$
(towards the RS2 phase),
the correlation between the regulated state and the
second eigenvector increases 
to be larger than
that between the first eigenvector and the regulated state.
Meanwhile,
as shown in \Fref{fig:corr_eig_and_m}(b),
the correlation between the 
first eigenvector and the non-regulated state $\bm{\mu}^-$
is always higher than that of other eigenvectors at $T<T_0$.
For the higher order-eigenvectors than the second-order,
their correlation between the regulated and non-regulated states
is small, as with $\bm{\xi}_3$ shown in Fig.\ref{fig:corr_eig_and_m}(a) and (b).

To summarize,
typical phenotypes evolved in the RS1 phase
are concentrated on the direction of the first eigenvector,
for both with and without regulation.
Meanwhile, in the RS2 phase,
the typical phenotypes with and without regulation are 
distinctively along the second 
eigenvector and the first eigenvector of genotype, respectively.
Thus, the typical phenotypes generated by
two distributions $p_\beta^\pm(\bm{S}|\bm{J})$
are almost orthogonal to each other.

The contributions of the first and second eigenmodes to the 
mean phenotypes $\bm{\mu}^{\pm}$ 
are given by the magnitudes of their 
corresponding eigenvalues.
In \Fref{fig:corr_eig_and_m}(c),
we show the $T$-dependence of the expected value of the
first, second, and third eigenvalues of the evolved genotypes.
The horizontal lines denote 
their expected values for the symmetric matrices
whose components independently and identically obey the uniform 
distribution over $\Omega_J$.
For $T<T_0$,
the first eigenvalue shows distinct increases from the expected value,
whereas for $T\simeq T_1$,
the second eigenvalue increases.
Meanwhile,
the third and higher-order eigenvalues show slight changes.
Therefore, 
the two desirable phenotypes
are achieved by the
contribution of the first and second order eigenmodes.

\begin{figure}
\begin{minipage}{0.495\hsize}
\centering
\includegraphics[width=2.7in]{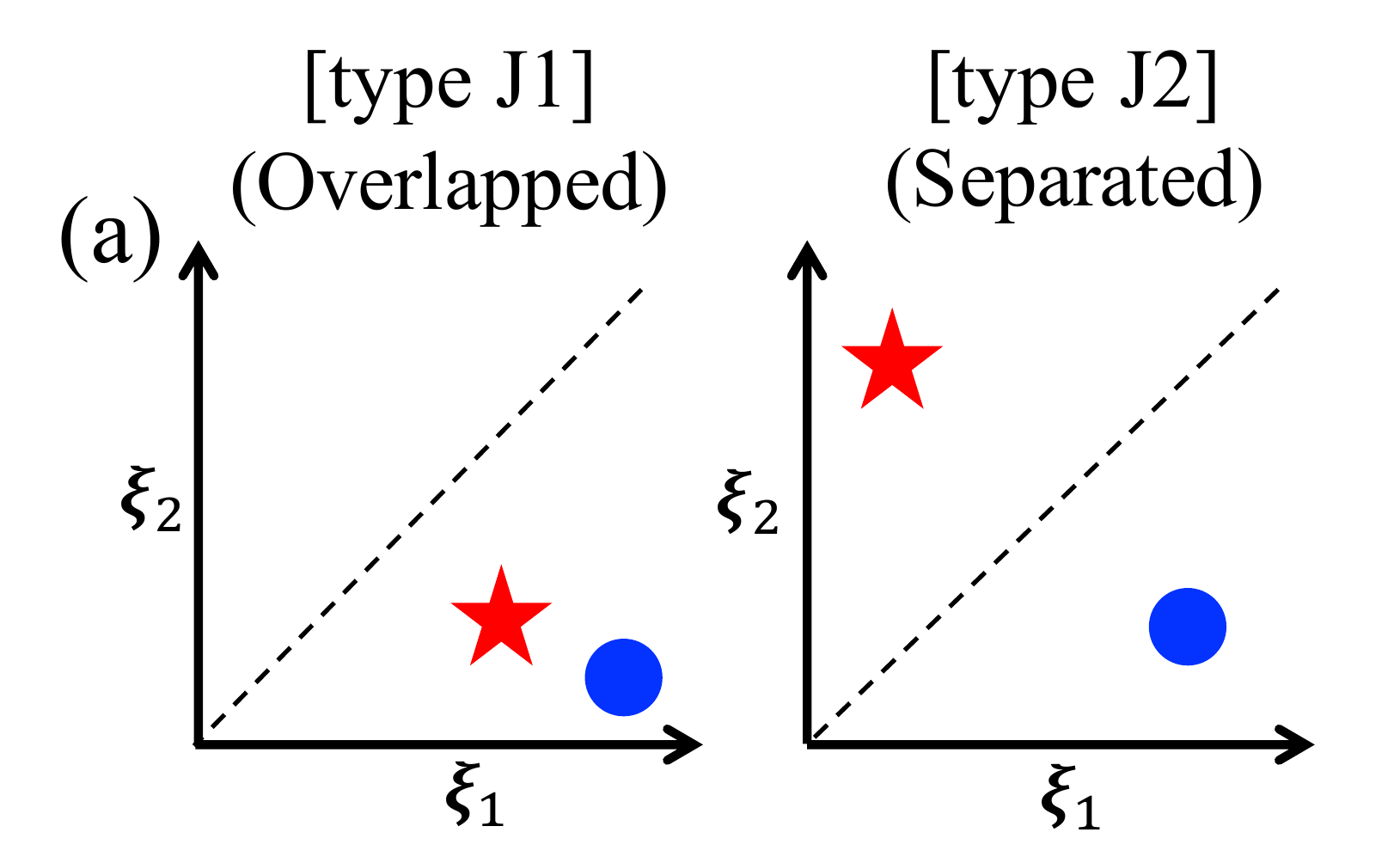}
\end{minipage}
\begin{minipage}{0.495\hsize}
\centering
\includegraphics[width=3in]{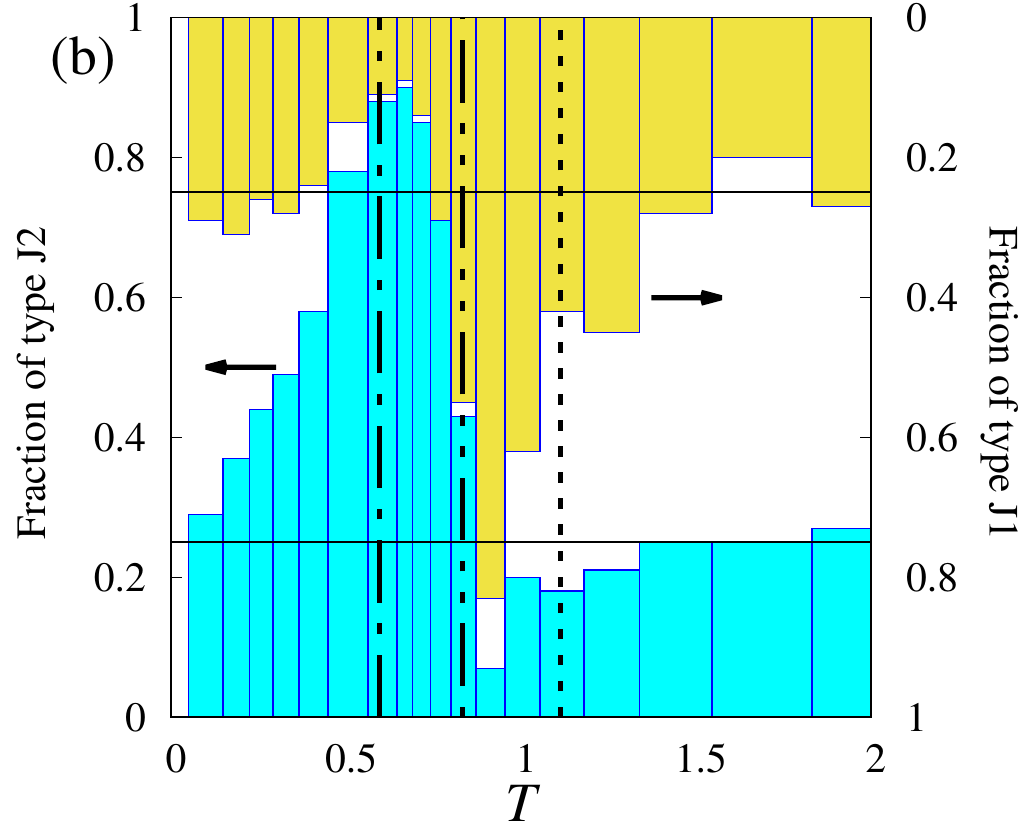}
\end{minipage}
\caption{
(a) Characteristic mapping of the mean phenotypes
with regulation ($\star$) and without regulation ($\bullet$),
where the 
diagonal dashed line with a 45-degree slope
is a guide for the eyes.
(b) Fraction of genotypes of
types J2 (cyan, left axis) and 
J1 (yellow, right axis).
The horizontal lines represent 0.25,
which is the trivial value for randomly distributed genotypes.
The vertical dashed line, one-dot chain, and two-dot chain line
represent $T_0$, $T_2$, and $T_1$, respectively.}
\label{fig:separable_fraction}
\end{figure}

Following these observations, 
we map the
mean phenotypes with and without
regulation onto the 
two-dimensional space spanned by
the first and second eigenvectors, $\bm{\xi}_1$ and 
$\bm{\xi}_2$, of the evolved genotypes.
In the RS1 and RS2 phases,
characteristic mappings 
are observed as shown in
Fig.\ref{fig:separable_fraction} (a),
where the mean phenotypes with and 
without regulation are denoted by $\star$ and $\bullet$, respectively.
In the RS1 phase, 
the first eigenvector
is dominant to express both 
mean phenotypes with and without regulation
for most of the evolved genotypes.
We term this case 
as the overlapped phenotypes 
(Fig.\ref{fig:separable_fraction} (a) left).
Meanwhile, in the RS2 phase,
phenotypes are shaped by the first 
and second eigenvectors
of the evolved genotypes. These genotypes can satisfy
the required fitness conditions
both without and with regulation, respectively.
We term the case as 
separated phenotypes 
as shown in
Fig.\ref{fig:separable_fraction} (a) right.
Hereafter, we term the genotype $\bm{J}$ that gives overlapped 
and separable phenotypes
as type J1 and type J2, respectively.

%

Fig.\ref{fig:separable_fraction}(b)
shows the temperature dependence of the fraction of the type J1
and J2 genotypes among the ensemble of evolved genotypes ${\cal J}(T)$.
At sufficiently large $T$, their 
fractions are equal to 0.25, which is indicated by horizontal lines.
The value of 0.25 is the expected value of the fraction of type J1 and J2
for the randomly generated $\bm{J}$s,
as there are two other cases of mapping;
the case that $\star$ and $\bullet$ located along $\bm{\xi}_2$,
and that $\star$ and $\bullet$ are along $\bm{\xi}_1$ and $\bm{\xi}_2$,
respectively.
As $T$ decreased toward the RS1 phase,
the fraction of genotypes of type J1
increases up to 0.8.
By lowering the temperature further in the RS2 phase,
the dominant genotype is replaced by type J2.
For lower $T<T_1$, the dominancy of type J2 decreases
as $T$ decreases,
and the fraction of types J1 and J2 approaches 0.25.

\subsection{Why the separable phenotype appears at $T<T_1$?}

Here, we discuss why 
the type J1 and J2 genotypes are dominant 
at $T_1<T<T_0$ and $T_2<T<T_1$, respectively.
To answer this question,
we observe the fitness of the evolved genotypes ${\cal J}(T)$
under a trial temperature $T_{\mathrm{tr}}$.
The evolutionary process in our model selected genotypes
among possible $\bm{J}$s;
hence, $\bm{J}\in{\cal J}(T)$ can be a candidate 
for genotypes in ${\cal J}(T_{\mathrm{tr}})$ ($T\neq T_{\mathrm{tr}}$), in principle.
By evaluating the fitness of $\bm{J}\in{\cal J}(T)$
at a different temperature $T_{\mathrm{tr}}$,
we discuss the reason why $\bm{J}\in{\cal J}(T)$
cannot be selected at different temperatures.

\begin{figure}
\centering
\includegraphics[width=3in]{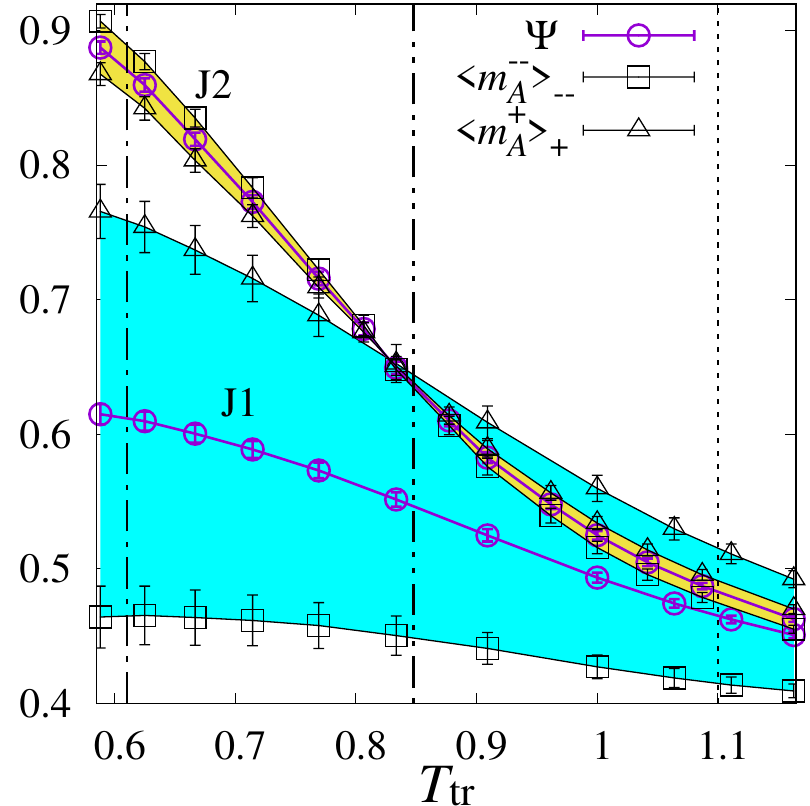}
\caption{Trial temperature $T_{\mathrm{tr}}$ dependence of the 
fitness for $\bm{J}\in{\cal J}(T)$
at $T=0.91$ (denoted by J1) and 
$T=0.63$ (denoted by J2).
The shaded region indicates the difference between
$\langle m_A^-\rangle_{-}$
and $\langle m_A^+\rangle_+$.}
\label{fig:different_temp}
\end{figure}

Fig.\ref{fig:different_temp}
shows the $T_{\mathrm{tr}}$ dependence of the fitness $\Psi$ 
on $\bm{J}\in{\cal J}(T)$ for $T=0.91$ (type J1; RS1)
and $T=0.63$ (type J2; RS2).
At sufficiently large $T_{\mathrm{tr}}$, J1 and J2 fitness values are
not much different.
Type J1 and J2 are subject to one- and two-dimensional constraints,
respectively, hence, the possible configurations of type J1
are larger than that of type J2.
From the thermodynamic perspective,
the dominance of the type J1 in the RS1 phase is caused by 
this entropic effect.
In the RS2 phase, the fitness of type J2 
is sufficiently large to overcome the entropic effect,
and they can be dominant in this phase.
This observation indicates that the 
changes in the ensemble of ${\cal J}(T)$ can be regarded as 
a phase transition with respect to genotypes
between the type J1 and J2.

\subsection{Evolutional dynamics of genotypes on the two-dimensional plane}

\begin{figure}
\centering
\includegraphics[width=5.2in]{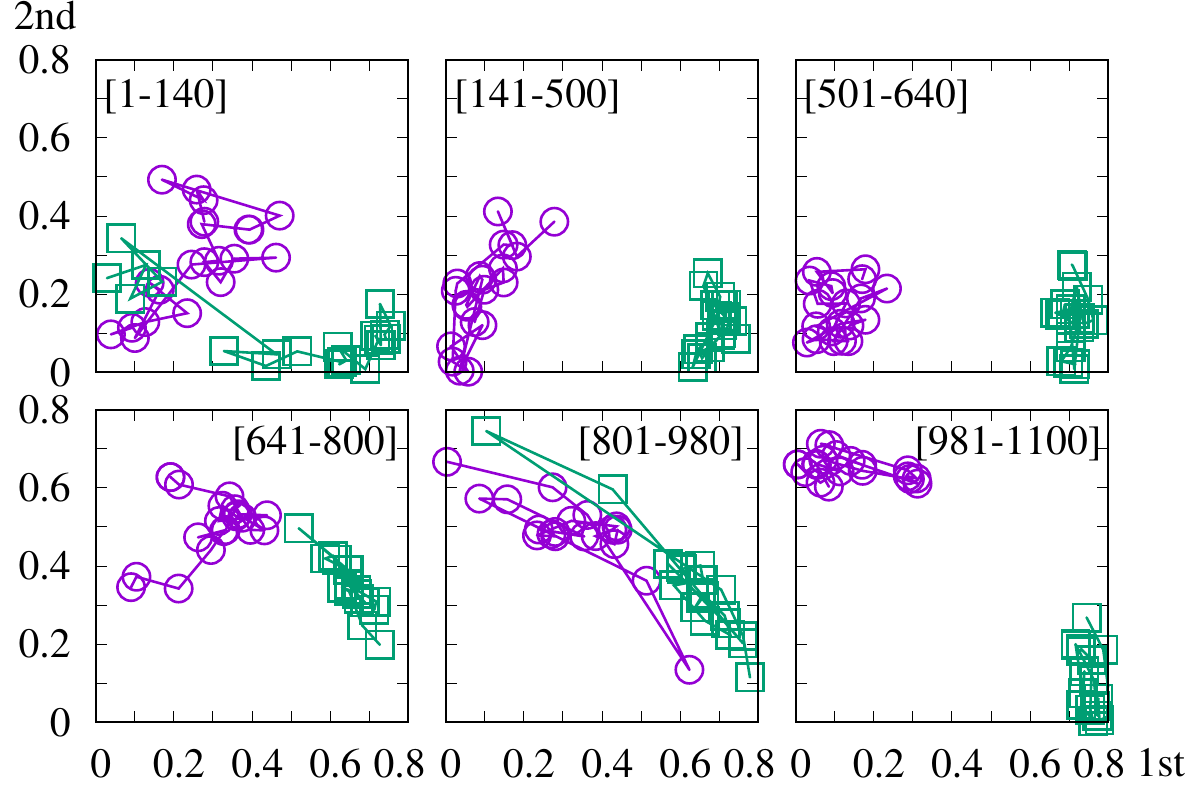}
\caption{The evolution of the mean phenotypes.
Corresponding to the evolution of Fig.\ref{fig:m_vs_gen} (b),
this figure shows the evolution of the mean phenotype in the two-dimensional
space spanned by the first and second eigenvectors of 
the genotype for the evolution generations
$[1-140]$, $[141-500]$, $[501-640],\cdots,[981-1100]$.
The results with and without regulations are plotted by 
$\circ$ and $\square$,
respectively.
}
\label{fig:gen_on_2dim}
\end{figure}

For the understanding of the evolutionary construction of the 
separated phenotypes,
we simplify the evolutionary dynamics
using the two-dimensional space spanned by the first and second eigenvectors, 
although the two-dimensional approximation was not necessarily accurate 
in the early stages of evolution, even in the RS2 phase.
In Fig.\ref{fig:gen_on_2dim},
we show evolutionary change of the mean phenotypes
in RS2 phase corresponding to the series shown in
\Fref{fig:m_vs_gen} (b), where
$\langle m_A^-\rangle$ increased before $\langle m_A^+\rangle$.
The panels of Fig.\ref{fig:gen_on_2dim} 
show the time evolution of the mean phenotypes
$\bm{\mu}^+(\bm{J}^{(g)})$ ($\bigcirc$) and $\bm{\mu}^-(\bm{J}^{(g)})$ ($\square$)
mapped onto the 
two-dimensional space spanned by the first and second eigenvectors
of the genotype at each generation denoted in the panels.
The localization of the mean phenotype without regulation
appears on the first eigenvector 
141-500 generations before that of the regulation case.
From generations 641-800, 
the contribution of the second eigenvector to the mean phenotype increases with regulation. 
After the reorganization of the distributions at generations 801-980,
the characteristic phenotype mapping for the type J2 genotype appeares.

When $\langle |m_A^+|\rangle_+$ increases before 
the increase of $\langle |m_A^+|\rangle_-$, 
the localization of $\bm{\mu}^+$ on the second eigenvector
appears in the early stage of evolution.
Additionally, the localization of $\bm{\mu}^-$ follows
with the reorganization of 
$\bm{\mu}^+$.
When both $\langle |m_A^\pm|\rangle_\pm$ increased simultaneously,
$\bm{\mu}^\pm$ are localized almost simultaneously (see supplement material).

\section{Switching trajectory}

\begin{figure}
\begin{minipage}{0.495\hsize}
\centering
\includegraphics[width=2.5in]{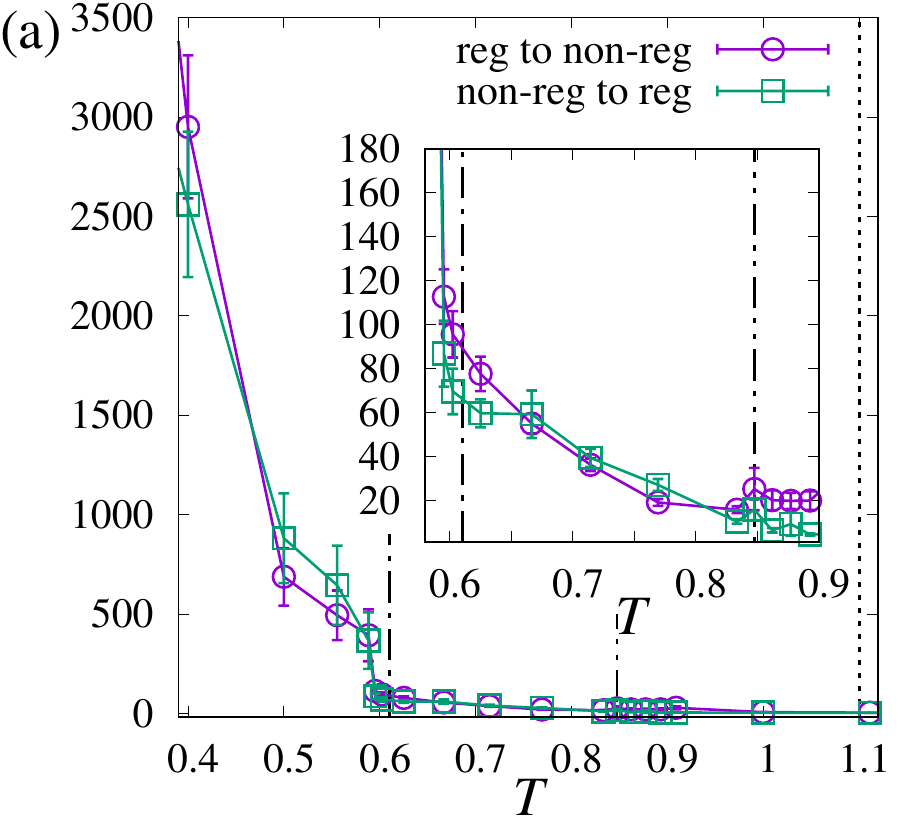}
\end{minipage}
\begin{minipage}{0.495\hsize}
\centering
\includegraphics[width=3in]{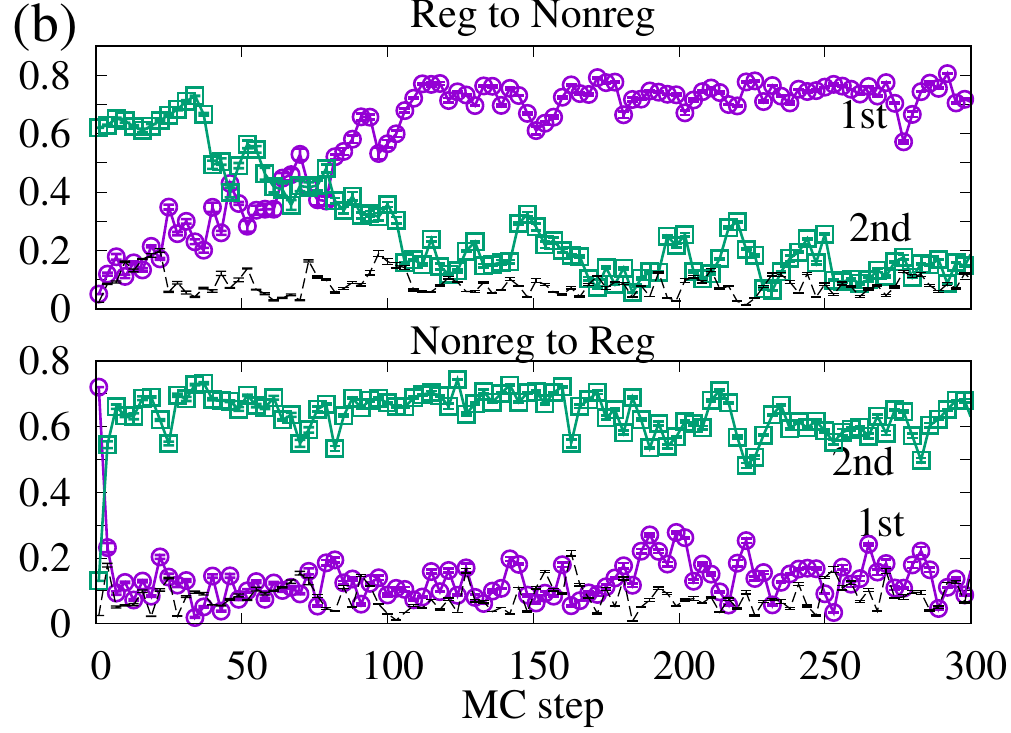}
\end{minipage}
\caption{(a) MC steps required for the switching from 
regulated to non-regulated state ($\bigcirc$) and 
non-regulated to regulated state ($\square$).
The dashed vertical line, two-dotted chain line, and 
one-dotted chain line denote $T_0$, $T_1$, and $T_2$, respectively.
The inset magnifies the difference between the RS1 and RS2 phase.
The transition time to shift the active sites
from the state without regulation to
that with regulation is evaluated as follows.
After the sufficient time updates of $\bm{S}$ under the non-regulated condition 
$\bm{S}_{\cal R}\notin \bm{S}_{\cal R}^+$,
the regulatory sites are changed to 
$\bm{S}_{\cal R}\in\bm{S}_{\cal R}^+$, 
and then $\bm{S}$ (except the regulatory region) is updated
according to (\ref{eq:S_expression}).
We compute
the target magnetization 
$|\sum_{i\in{\cal A}}S_i/N_A|$ at each MC step
to obtain the step where $|\sum_{i\in{\cal A}}S_i/N_A|$ first reaches 
the value $\langle|m_A^+|\rangle_+$,
which is defined as the transition time.
(b) Switching trajectories of 
local magnetizations projected to the first and the second eigenvectors
defined on an evolved genotype at $T=0.67$ (RS2).
The component projected onto the 
third order eigenvector is denoted by dashed lines.}
\label{fig:switch_time}
\end{figure}

\begin{figure}
\centering
\includegraphics[width=3in]{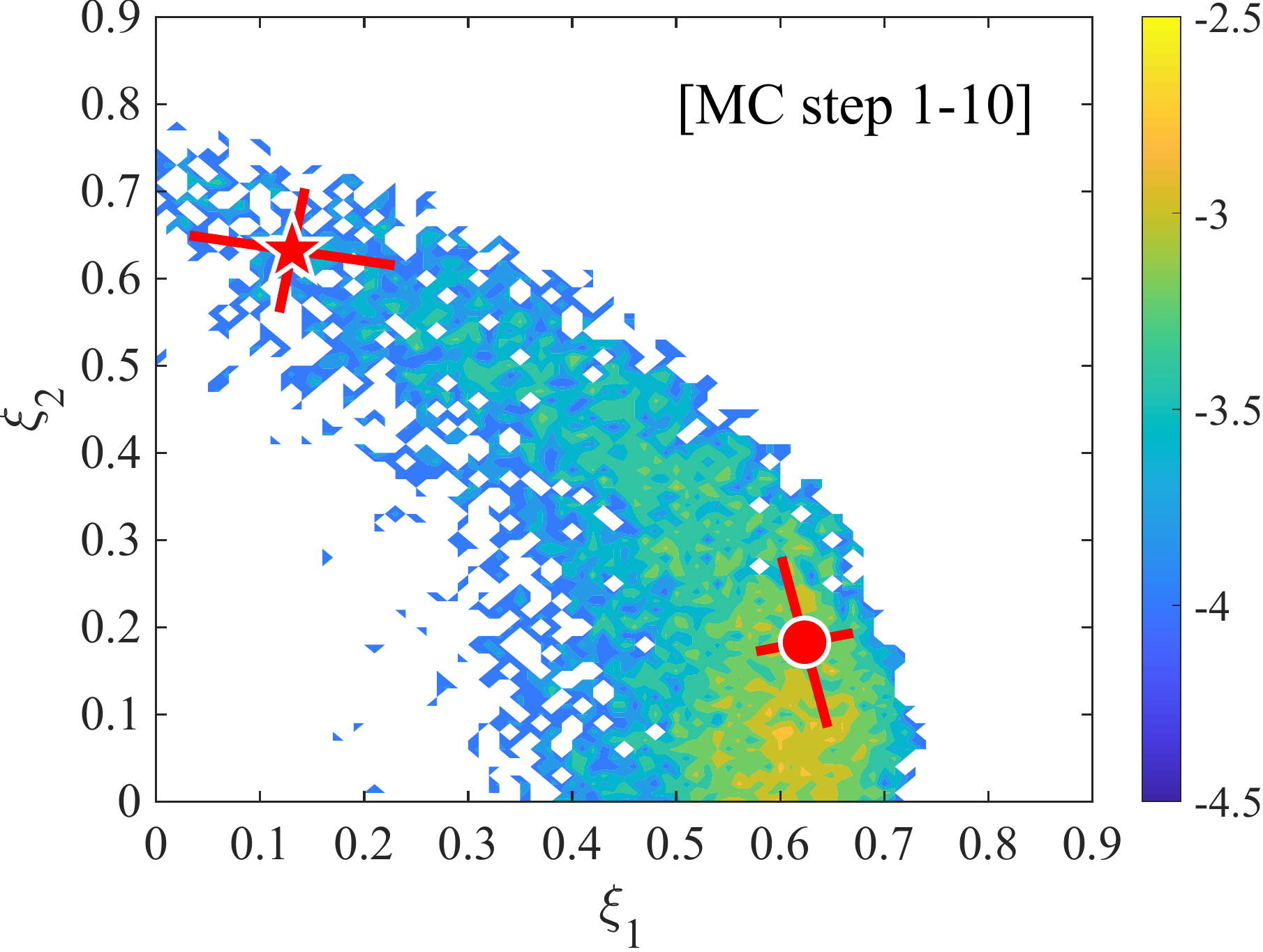}
\includegraphics[width=3in]{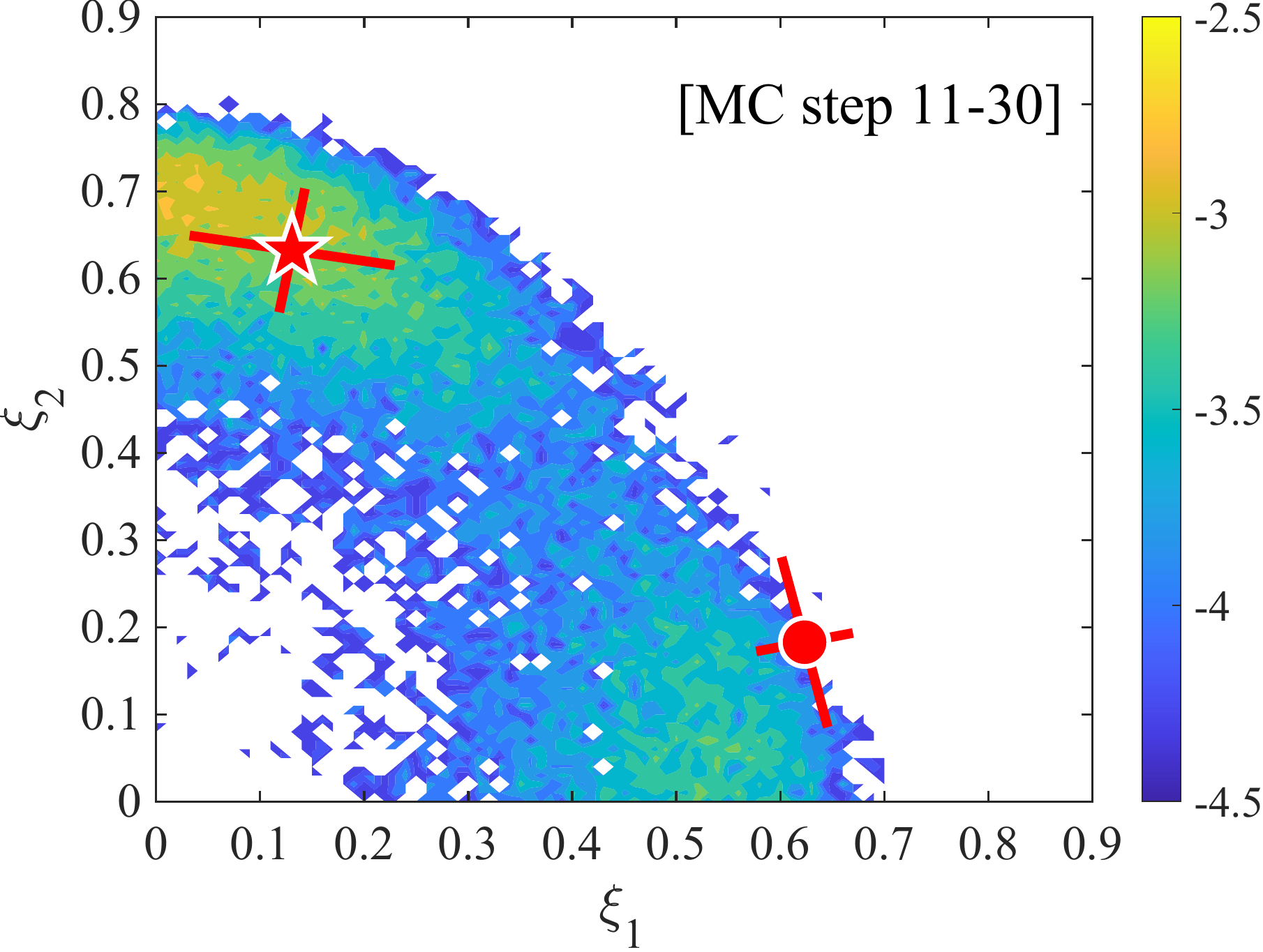}
\includegraphics[width=3in]{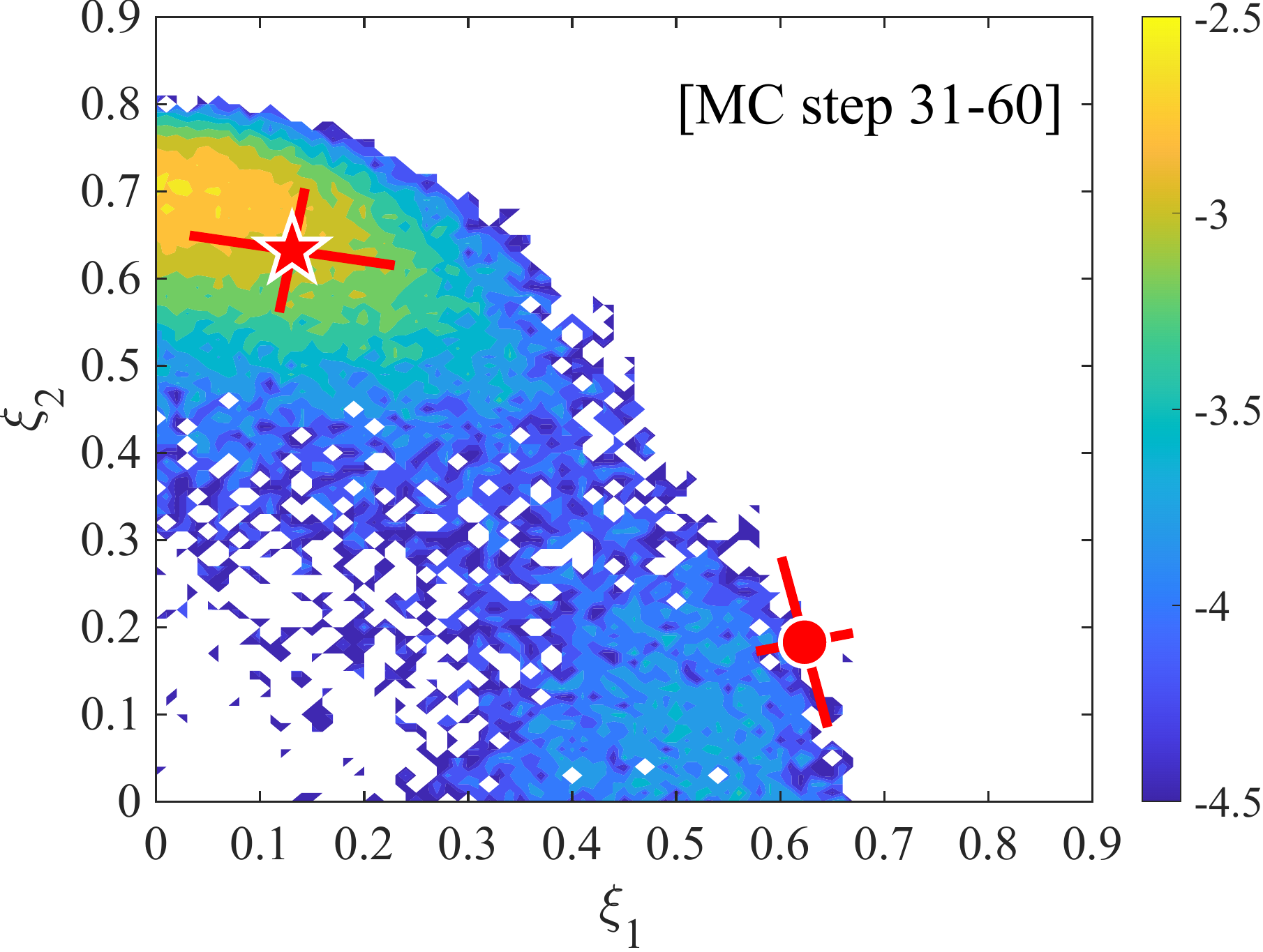}
\includegraphics[width=3in]{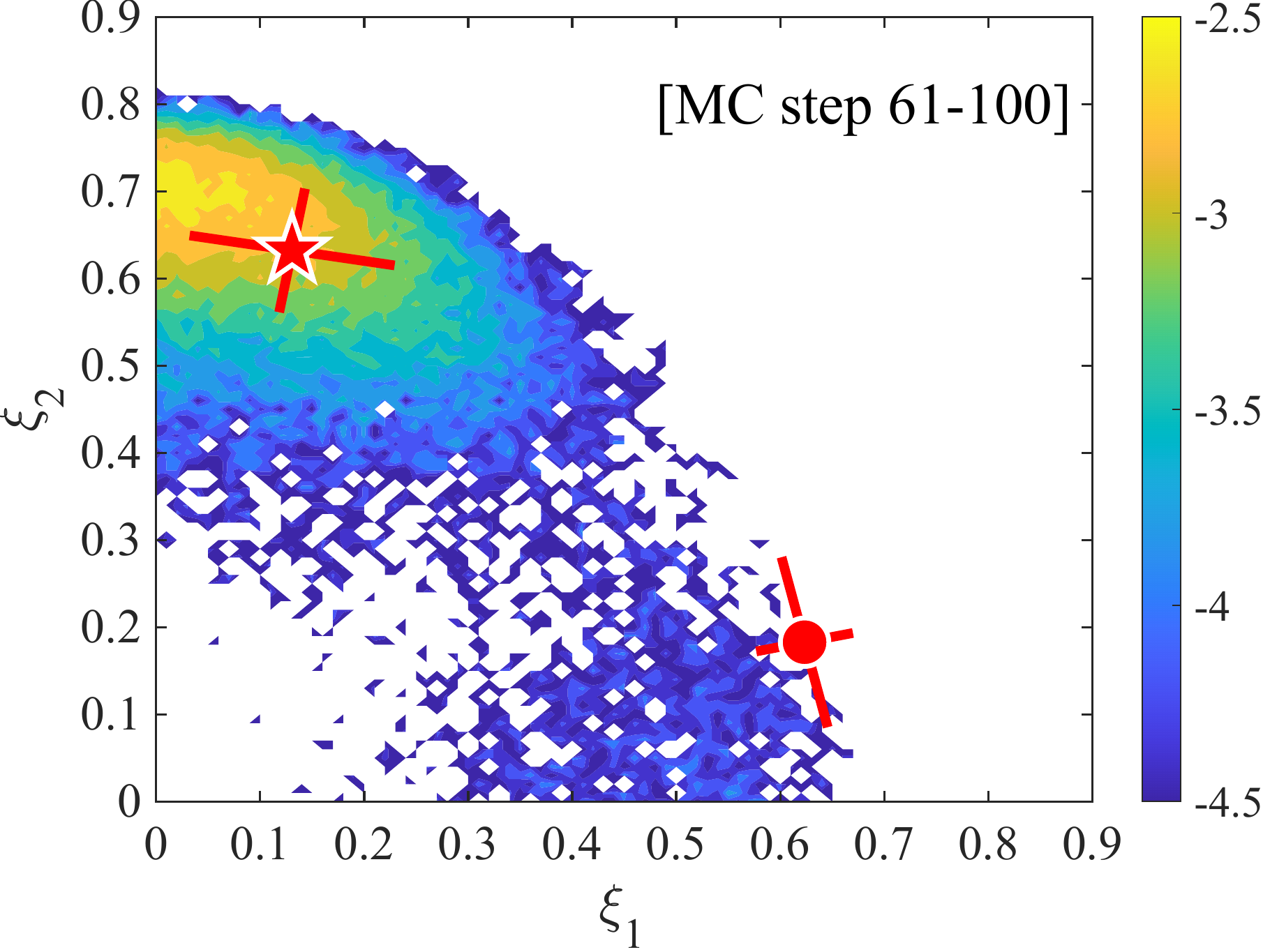}
\caption{Heat maps on the two-dimensional space
for switching trajectories
from regulated state to non-regulated state,
defined on an evolved genotype at $T=0.67$ (RS2).
Here, the two-dimensional space is meshed by 0.01,
and $\log_{10}$-frequencies of the trajectories 
during the given steps are plotted.
$\star$ and $\bullet$ denote the regulated state and 
non-regulated state projected onto the two-dimensional space, respectively.
The lines below these points represent the first and second eigenmodes
of fluctuation around these points.
Here, the length of the lines is magnified to be discernible. 
However, the ratio of the lines 
is proportional to the root of the ratio of the eigenvalues.}
\label{fig:switch_trajectory_nonreg2reg}
\end{figure}


Under the $\bm{J}$s of type J2, 
the shift between the regulated and non-regulated states
involves a large conformational change.
We employ the MCMC method according to (\ref{eq:S_expression})
for simulating the transition dynamics 
from regulated to non-regulated cases,
or from non-regulated to regulated cases,
and compute the MC steps required for the shift between the two cases.

Fig. \ref{fig:switch_time}(a) shows the 
transition time calculated by the MCMC method
from the regulated
to non-regulated states ($\bigcirc$),
and from the regulated to non-regulated states ($\square$),
respectively.
Here, the upper limit of the MC step is set at $10^5$.
In the RS1 phase, 
there is little change in phenotypes with and without regulation, and 
the transition time is 
within 20 steps.
Compared with the RS1 phase,
the transition time required in the RS2 phase increases.
This increase in relaxation time is associated with 
the large conformational change of phenotype under the J2 genotypes.
However, the large conformational 
change does not qualitatively change the relaxation time. 
As in the RS1 phase, 
the relaxation time in the RS2 phase is in the order of $10^2$.
In the RSB phase, the MC steps required for switching diverge
as $T$ decreases.
This phenomenon in the RSB phase
is consistent with the property of the RSB phase
where the metastable states hamper relaxation.

The trajectories shifting between two states 
lie in $2^N$-dimensional space.
However, particularly in the RS2 phase, 
the two-dimensional space spanned by
the first and second eigenvectors of the evolved genotype
is sufficient to 
describe the switching trajectories.
This 
low-dimensional constraint is already observed 
as 
the equilibrium property in the RS2 phase, 
as shown in Fig.\ref{fig:corr_eig_and_m} (a) and (b).
Fig. \ref{fig:switch_time} (b) shows 
the trajectories of the components
projected onto the first ($\circ$),
second ($\square$), and third (dashed line) eigenvectors
defined on an evolved $\bm{J}$ of type J2 at $T=0.67$ (RS2).
In the regulate-to-non-regulate switching,
the change in the first component is much larger,
and in the non-regulate-to-regulate switching,
the change in the second component is much larger.
Meanwhile, 
the third-order (and higher) components are nearly constant 
during regulated-to-non-regulated or non-regulated-to-regulated switching.


We generate 1000 switching trajectories on
a certain $\bm{J}\in{\cal J}(T)$,
and map them onto the two-dimensional space
spanned by the first and second eigenvectors of the evolved genotypes.
Fig.\ref{fig:switch_trajectory_nonreg2reg}
shows the heat map on the two-dimensional space for the switching trajectories 
defined on an evolved genotype of type J2 at $T=0.68$ (RS2) from 
non-regulated to regulated states.
The mean phenotypes with and without regulation, $\bm{\mu}^+$ and $\bm{\mu}^-$,
after sufficient time steps of updating are
denoted by $\star$ and $\bullet$, respectively.
Additionally, the direction of the fluctuation of these points is indicated by two lines 
below the points.
The switching trajectory when regulation is removed is shown in the supplementary material.
For both cases of switching,
most of the trajectories 
follow a quarter-circle path.
%
%
%
This quarter-circle path
is restricted to a one-dimensional path
within the two-dimensional space.
With this restriction,
the transition time between the two states 
remains small, even though the two phenotypes are far apart,
as shown in Fig. \ref{fig:switch_time}(a).

\begin{figure}
\centering
\includegraphics[width=3in]{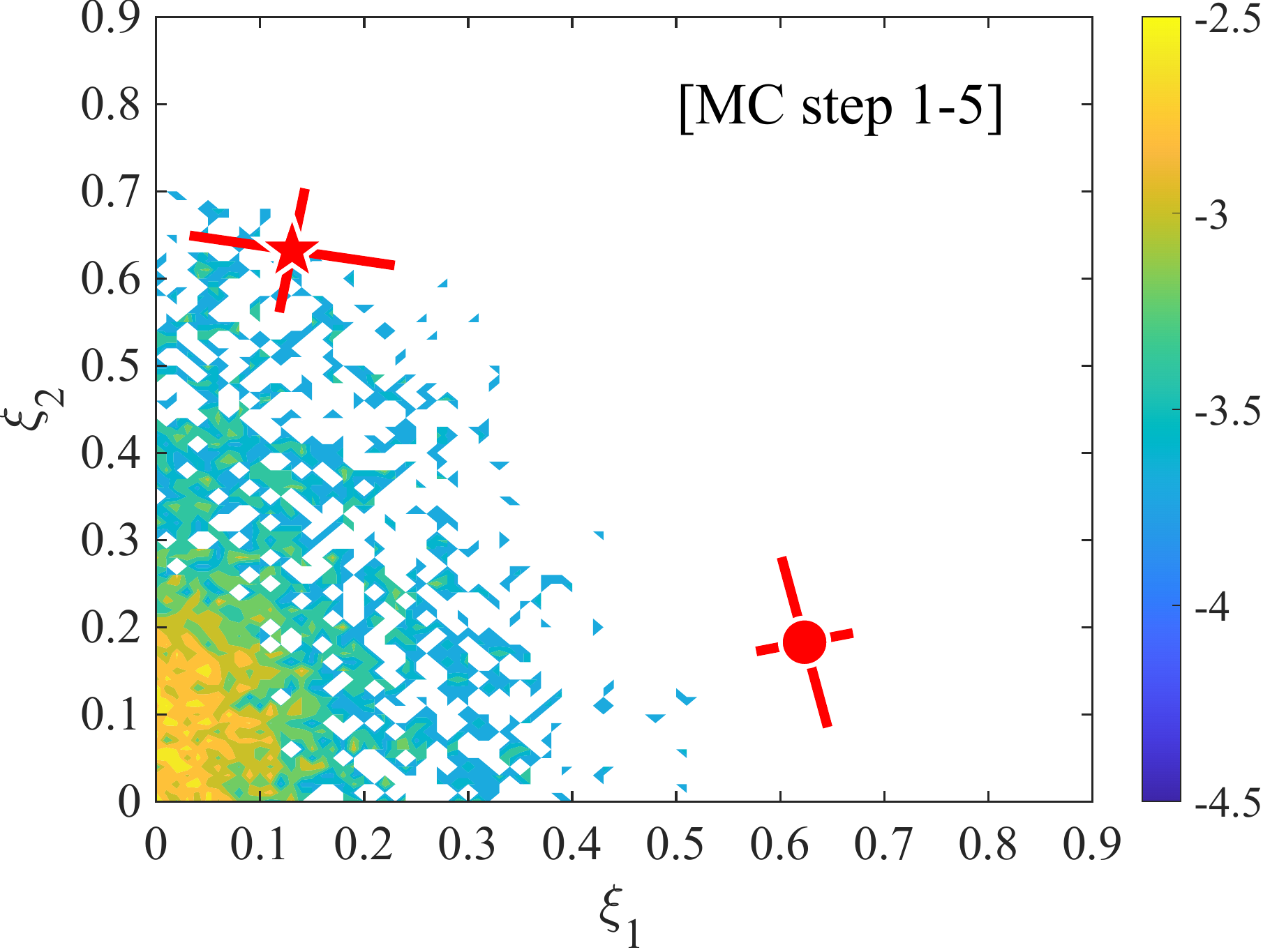}
\includegraphics[width=3in]{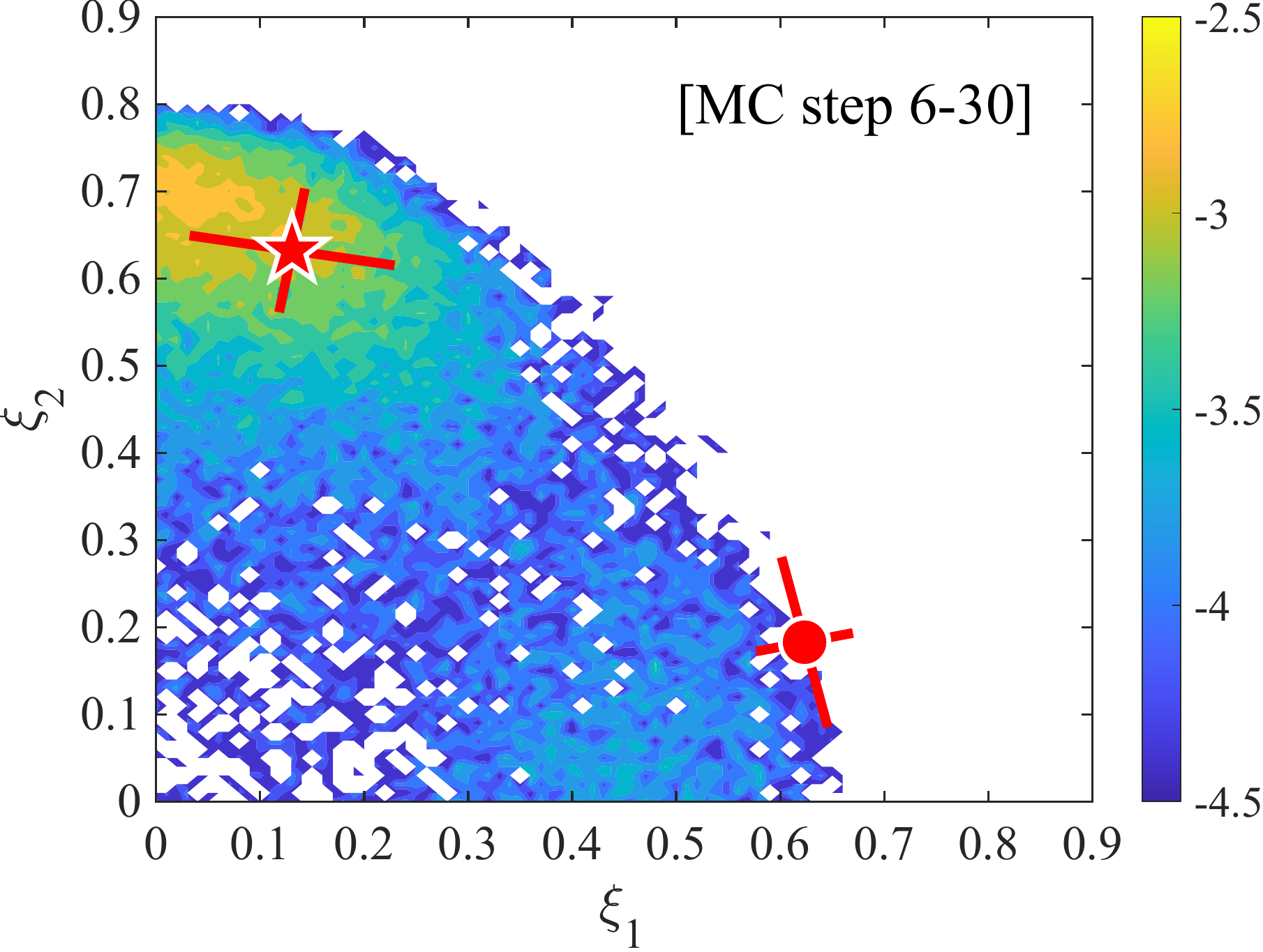}
\includegraphics[width=3in]{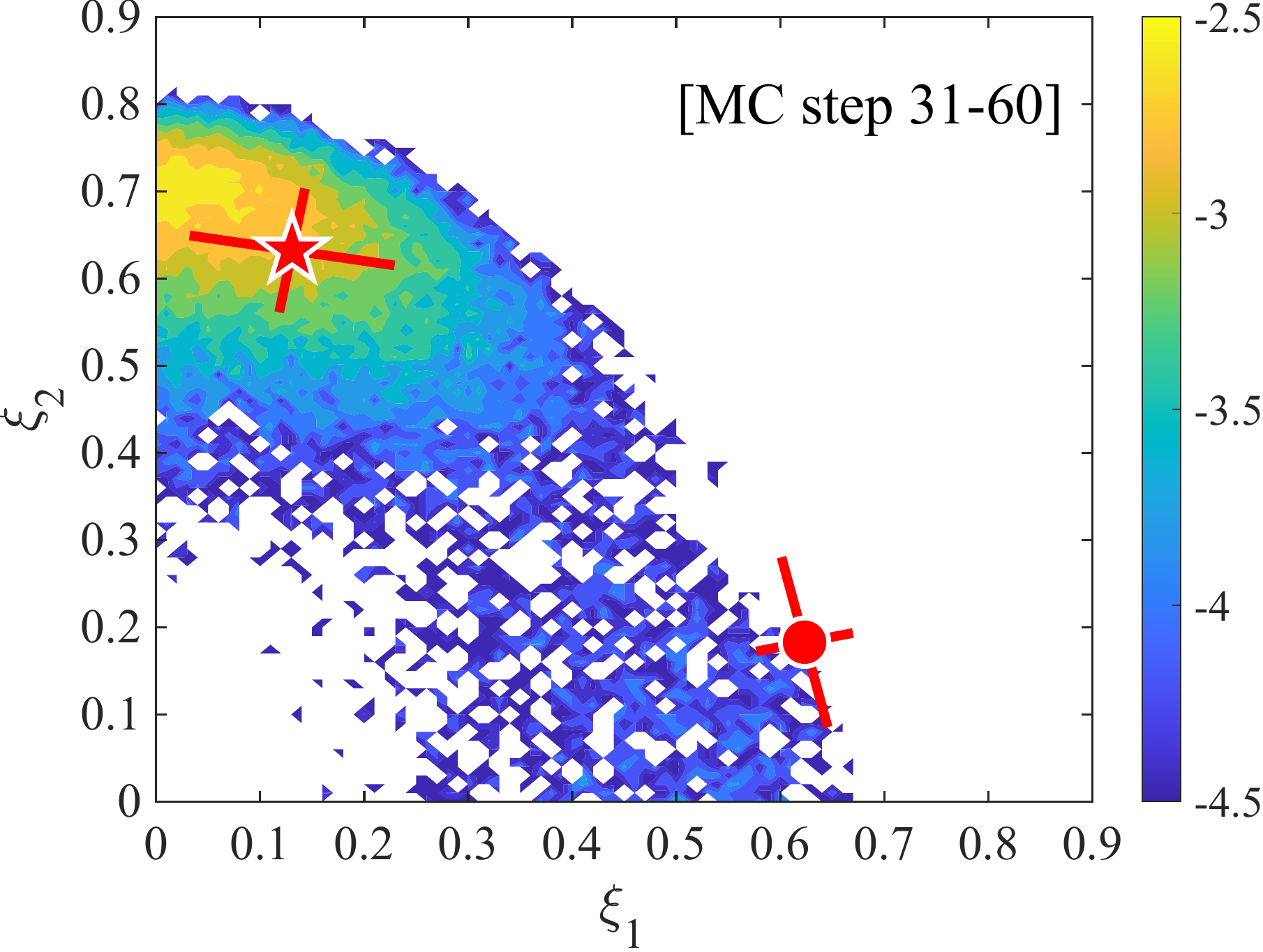}
\includegraphics[width=3in]{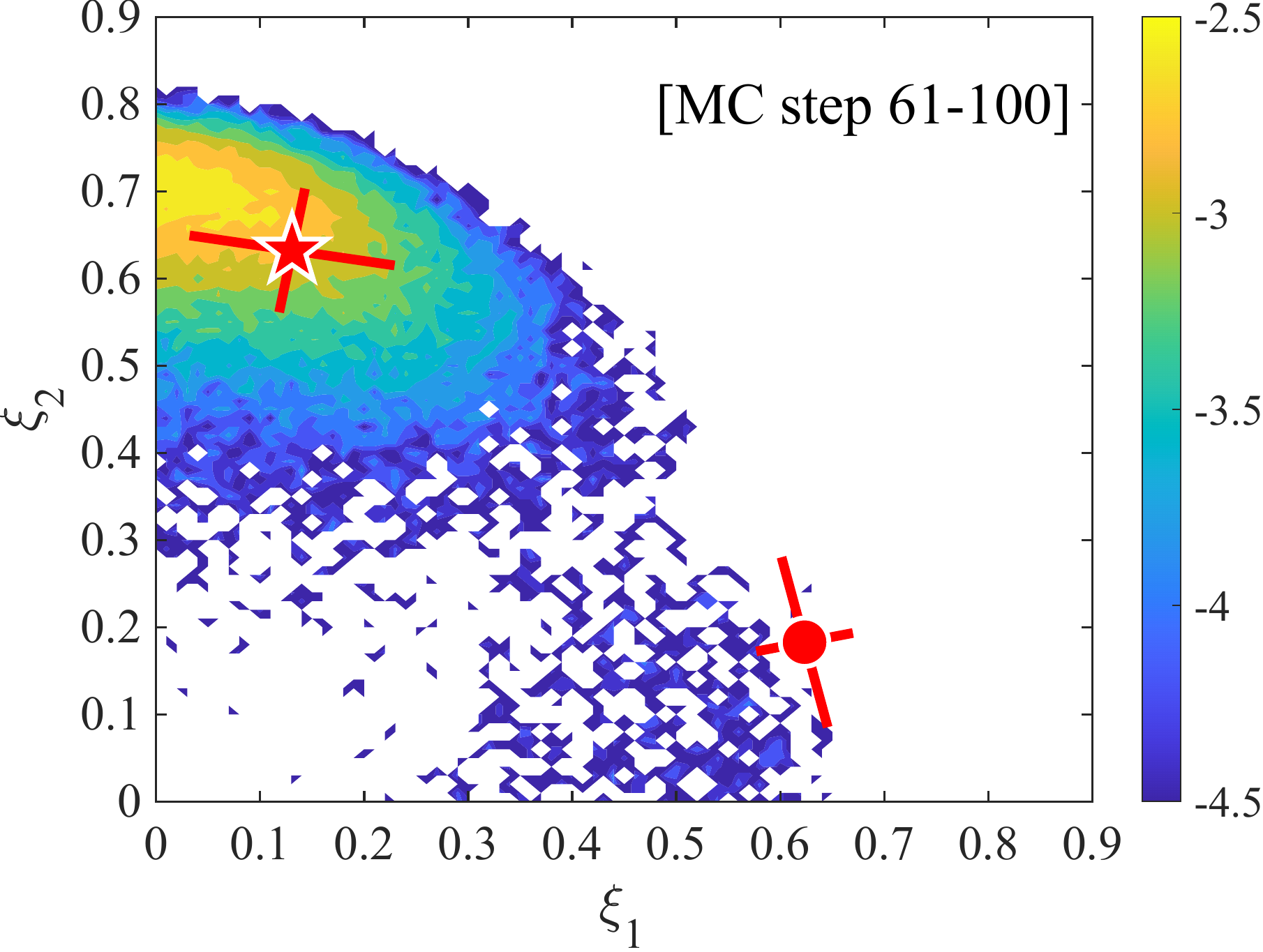}
\caption{Heat maps on the two-dimensional space
for relaxation trajectories from an initial condition 
to the regulated state
defined on the evolved genotype of type J2 at $T=0.67$ (RS2),
the same value as adapted in Fig.\ref{fig:switch_trajectory_nonreg2reg}.
$\star$, $\bullet$,
and the orthogonal lines below these points 
are the same as in Figs.\ref{fig:switch_trajectory_nonreg2reg}.
}
\label{fig:switch_trajectory_random2reg}
\end{figure}

The quarter-circle path on the two-dimensional plane
restricts the trajectories of the convergence from arbitrary initial conditions
to the phenotypes with and without regulation.
The heat map for the relaxation dynamics 
on a type J2 genotype evolved at $T=0.67$ from 
arbitrary initial conditions 
is shown in Fig. \ref{fig:switch_trajectory_random2reg}
for the regulated case.
Most of the trajectories 
are attracted once to the quarter-circle line
where the switching paths are concentrated,
and then approach the regulated state.
(The relaxation dynamics of the non-regulated phenotype
are shown in the supplementary material.)
The quarter-circle path is attractive in the sense that 
any state tends towards the regulated or non-regulated state 
through this path.

\section{Two dimensional approximation of free energy landscape}

To understand the characteristic switching path in the 
two-dimensional space,
we examine the free energy landscape.
The free energies for the regulated and non-regulated cases,
denoted by $f_+$ and $f_-$, are defined as
\begin{align}
f_+&=-\frac{1}{N\beta}\ln\sum_{\bm{S}|\bm{S}_{{\cal R}}\in\bm{S}_{{\cal R}}^+}\exp(-\beta H), \\
f_-&=-\frac{1}{N\beta}\ln\sum_{\bm{S}|\bm{S}_{{\cal R}}\notin\bm{S}_{{\cal R}}^+}\exp(-\beta H).
\end{align}

Following the result of the numerical simulations,
we consider the two-rank approximation of the evolved $\bm{J}$ as 
$\bm{J}\simeq\lambda_1\bm{\xi}^1{\bm{\xi}^1}^{\top}
+\lambda_2\bm{\xi}^2{\bm{\xi}^2}^{\top}$.
Under the two-rank approximation, the Hamiltonian is given by
\begin{align}
H=-\sum_{k=1}^2\frac{\lambda_k}{2}\left\{\left(\sum_{i=1}^N\xi_i^kS_i\right)^2-1\right\}.
\end{align}
For the two-rank approximation form, 
one can represent the free energy as a function of $m_1$ and $m_2$
defined by
\begin{align}
m_1^\pm &= \frac{1}{\sqrt{N}}{\bm{\xi}^1}^\top\bm{\mu}^\pm,\\
m_2^\pm &= \frac{1}{\sqrt{N}}{\bm{\xi}^2}^\top\bm{\mu}^\pm,
\end{align}
where $m_1^+$ and $m_2^+$ correspond to the projection of the 
local magnetization with regulation onto
the first and second eigenvectors, respectively,
whereas $m_i^-~(i=1,2)$ are those without regulation. 
Following the calculation shown in the Appendix,
the free energies are given by
\begin{align}
f_+&=\!\!\sum_{k=1}^2\frac{\lambda_k{m_k^+}^2}{2}
-\frac{1}{N\beta}\left\{\sum_{i=1}^N\ln\!\left(2\cosh(\beta h_i^+)\right)\!
+\!\log(p_+^++p_-^+)\!\right\}+\frac{1}{N}\sum_{i<j, i,j,\in{\cal R}}J_{ij}
\label{eq:f_plus}\\
f_-&=\!\sum_{k=1}^2\frac{\lambda_k{m_k^-}^2}{2}
-\frac{1}{N\beta}\left\{\sum_{i=1}^N\ln\left(2\cosh(\beta h_i^-)\right)+\ln\left(1-\left(p_+^-+p_-^-\right)\right)\right\}
\label{eq:f_minus}
\end{align}
where $h_i^\pm=\lambda_1m_1^\pm\sqrt{N}\xi_i^1+\lambda_2m_2^\pm\sqrt{N}\xi_i^2$
and 
\begin{align}
p_\pm^+=\prod_{i\in{\cal R}}\frac{\exp(\pm\beta h_i^+)}{2\cosh(\beta h_i^+)},~~~
p_{\pm}^-=\prod_{i\in{\cal R}}\frac{\exp(\pm\beta h_i^-)}{2\cosh(\beta h_i^-)}.
\end{align}
The saddle point equations for $m_k^\pm$ are given by
\begin{align}
m_k^+&=\frac{1}{\sqrt{N}}\sum_{i\notin{\cal R}}\xi_i^k\tanh(\beta h_i^+)
+\frac{1}{N}\sum_{j\in{\cal R}}\xi_j^k\frac{p_+^+-p_-^+}{p_+^++p_-^+}\\
m_k^-&=\frac{1}{\sqrt{N}}\sum_{i\notin{\cal R}}\xi_i^k\tanh(\beta h_i^-)+
\frac{1}{\sqrt{N}}\sum_{i\in{\cal R}}\xi_i^k\frac{\tanh(\beta h_i^-)-(p_+^--p_-^-)}{1-\left(p_+^-+p_-^-\right)}.
\end{align}

\begin{figure}
\centering
\includegraphics[width=3in]{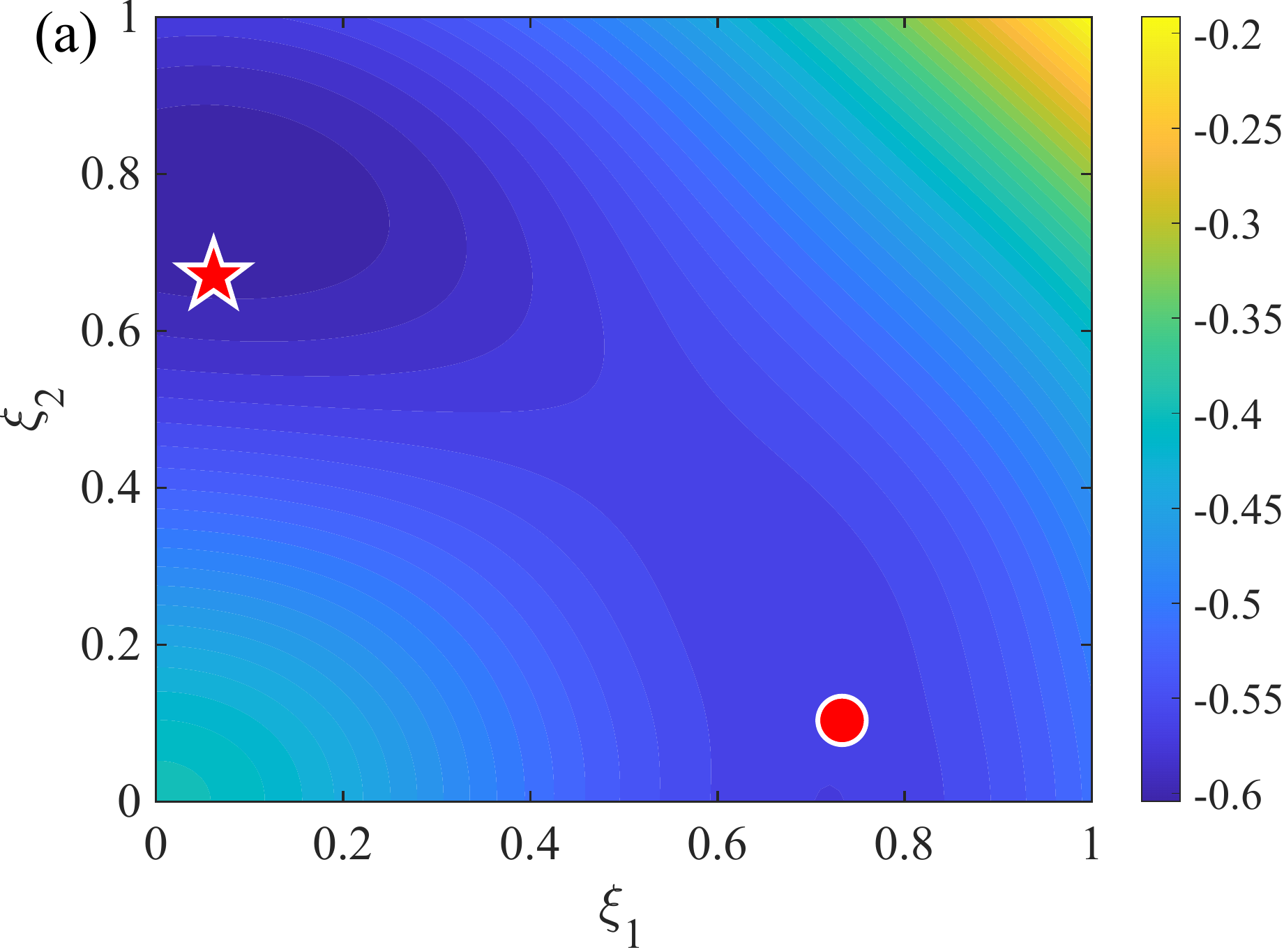}
\includegraphics[width=3.15in]{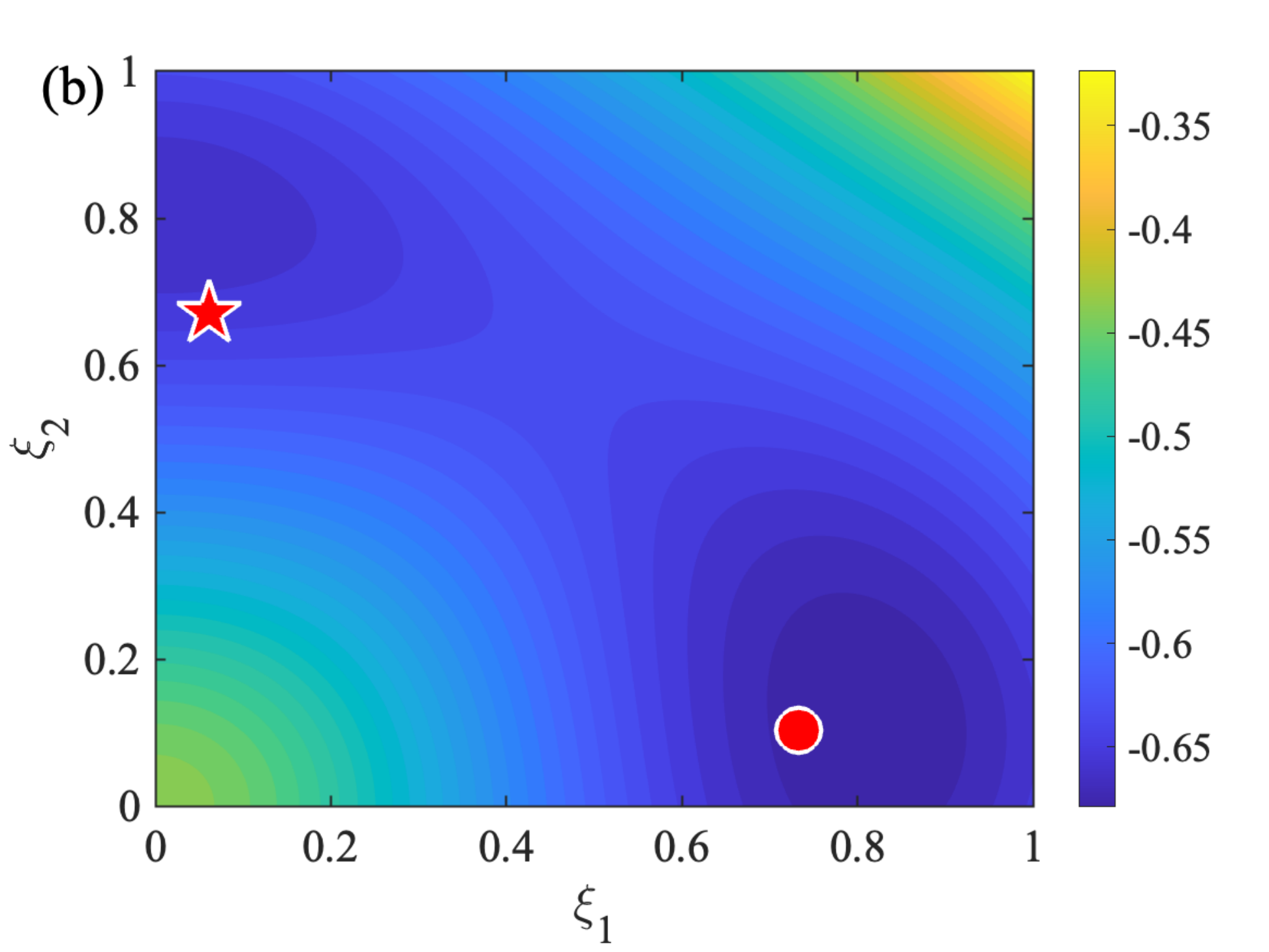}
\caption{Two-dimensional approximation of the free energy landscape
of one genotype with separable phenotype space
for (a) regulated case and (b) non-regulated case
at $p_{A}=0.05$ and $p_{R}=0.1$.
$\star$ and $\bullet$ show the projection of the regulated state and non-regulated state.}
\label{fig:free_energy_2dim}
\end{figure}

\Fref{fig:free_energy_2dim} (a) and (b) show
the landscape of 
$f^+$ and $f^-$, respectively, plotted on the two-dimensional space
of one evolved J2 genotype in the RS2 phase at $T=0.67$.
The minima of the free energies are consistent with the 
numerically observed phenotypes with and without regulation,
which are indicated by $\star$ and $\bullet$;
hence, the two-dimensional approximation of the free energy
is valid.
As shown in \Fref{fig:free_energy_2dim},
along the quarter-circle shape that connects the 
regulated and non-regulated states,
the free energy remains small.
The trajectories shown in 
Fig. \ref{fig:switch_trajectory_nonreg2reg} are restricted to this quarter-circle,
wherein free energy is small.


\subsection{Rough approximation of free energy}

What property of the evolved genotype 
gives the quarter-circle shape of the free energy landscape.
We introduce the following assumptions.
\begin{description}
\item [A1] The difference between the 
first and second eigenvalues
is negligible.
\item [A2] The components of the eigenvectors 
$\bm{\xi}^1$ and $\bm{\xi}^2$
are independently and identically distributed, 
according to the Gaussian distribution ${\cal N}(0,1\slash N)$.
\item [A3] The active and regulatory sites are negligible.
\end{description}
Following assumption {\bf A1}, we
replaced the first and second eigenvalues $\lambda_1$ and $\lambda_2$
with their mean $\overline{\lambda}$.
Under these assumptions,
we obtained the following form of free energy, as explained in the Appendix;
\begin{align}
f_{\mathrm{app}}=\frac{\overline{\lambda}}{2}(m_1^2+m_2^2)
-\frac{1}{\beta}\int Dz\ln\cosh\left(\beta\overline{\lambda}\sqrt{m_1^2+m_2^2}z\right),
\label{eq:free_energy_app}
\end{align}
where $Dz=\frac{dz}{\sqrt{2\pi}}\exp\left(-\frac{z^2}{2}\right)$.
The form of (\ref{eq:free_energy_app})
indicates that the free energy under the approximations 
{\bf A1}-{\bf A3} depends on $m_1$ and $m_2$ through
$m_S\equiv \sqrt{m_1^2+m_2^2}$.
Hence, the approximated free energy has the same value 
according to $m_S$, even when the 
individual values of $m_1$ and $m_2$ are different.
The saddle point of $m_S$ is given by
\begin{align}
m_S=\int Dz~z\tanh(\beta\overline{\lambda}m_Sz).
\end{align}

\begin{figure}
\centering
\includegraphics[width=3in]{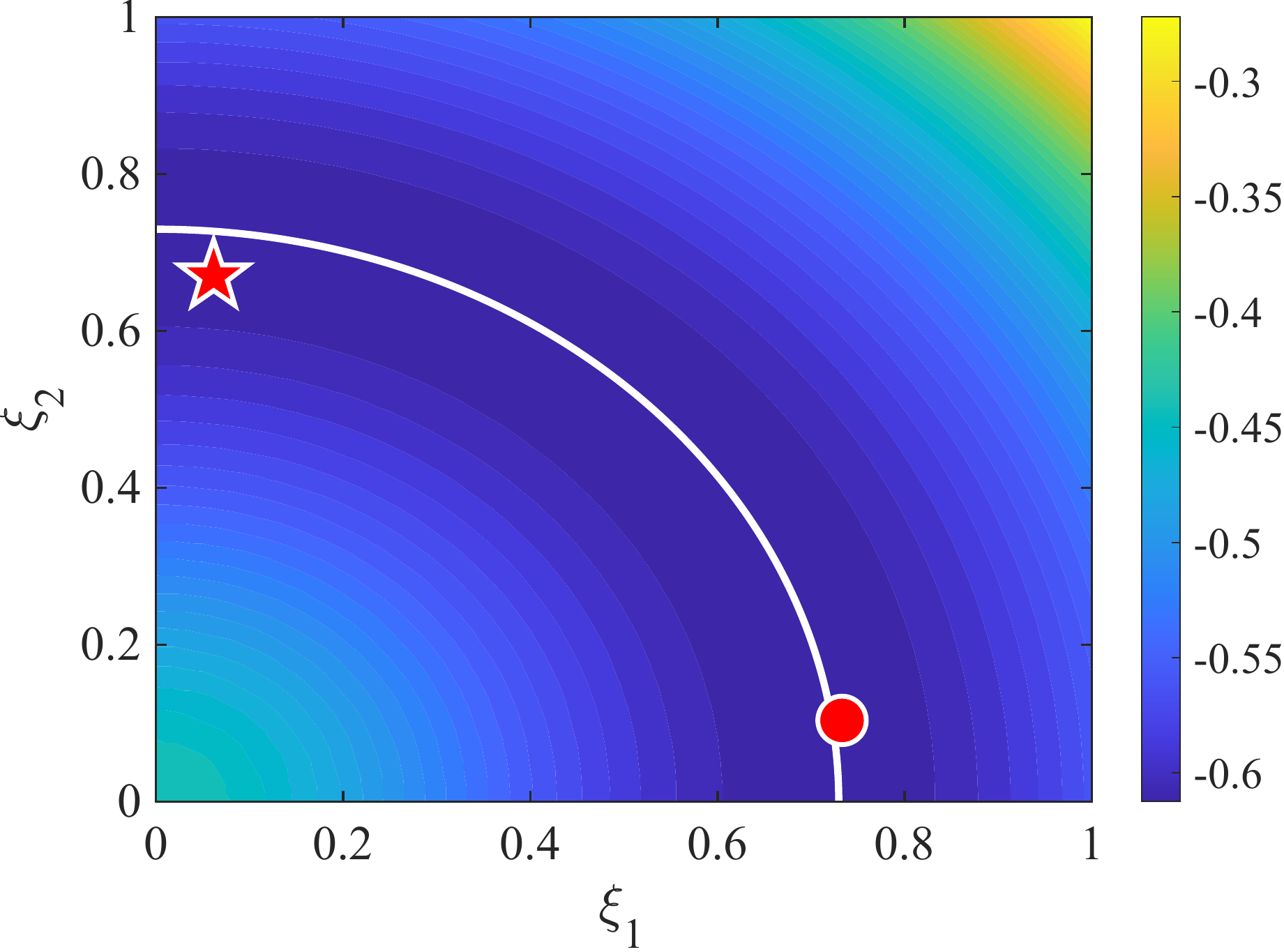}
\caption{
Free energy landscape on the two-dimensional plane
under the assumptions {\bf A1}-{\bf A3} at $T=0.67$ (RS2).
The solid line represents the minimum free energy
and the star and circle represent 
the mean of phenotypes with and without regulation, respectively.}
\label{fig:F_rough}
\end{figure}

Fig. \ref{fig:F_rough} shows the 
free energy landscape under the assumptions {\bf A1}-{\bf A3}
defined on a J2 genotype evolved at $T=0.67$ (RS2).
As expected from the form of eq. (\ref{eq:free_energy_app}),
the approximated free energy shows a quarter-circle landscape.
The quarter-circle curve
represents the minimum of the free energy $f_{\mathrm{app}}$,
whereas the equilibrium states with and without regulation
are denoted by stars and circles, respectively,
which are located near the extremum line of $f_{\mathrm{app}}$.
Therefore, the one-dimensional and quarter-circle switching path is 
considered to be provided by the free sites,
as the active and regulatory sites 
are ignored in deriving $f_{\mathrm{app}}$
(assumption {\bf A3}).
A particular difference between $f_\pm$ and
$f_{\mathrm{app}}$
is that the valleys around the 
mean of the phenotype with and without regulation (Fig. \ref{fig:free_energy_2dim})
cannot be described by $f_{\mathrm{app}}$.
For the description of these valleys, 
it is necessary
to consider
the active and regulatory sites.
Thus, free energy consists of 
quarter-circle switching paths provided by free sites
and valleys around the mean phenotypes
provided by active and regulatory sites.
Further, the assumption {\bf A2}
suggests that randomness in the embedded pattern in the free sites is 
significant for the description of the quarter-circle path.
Therefore, 
as the number of free sites decreases,
or equivalently as the number of active and regulatory sites increases,
the description by
$f_{\mathrm{app}}$ would be invalid.

\section{Summary and Conclusion}

In this study, we investigated the evolution of a spin model 
to generate two specific configurations of active sites
depending on the regulation. 
A fitness function was designed to increase 
when the appropriate spin configurations 
(phenotypes) with and without regulation 
appears with high probability. 
Our analysis revealed three transition points, $T_0$, $T_1$ and $T_2$: 
The fitness was increased from the trivial 
value for $T<T_0$. For $T_2<T<T_0$, 
the evolved system belonged to the RS phase. 
The RS phase was further divided into two regions at $T=T_1$, the RS1 
($T_1<T<T_0$) and RS2 ($T_2<T<T_1$) phases, 
with dominant genotypes differing in 
these regions, type J1 for RS1 and J2 for RS2 phases. 
For $T_1<T<T_0$, 
the phenotypes, i.e., spin configurations, other than active sites
barely depended on the regulation.
In contrast, for $T_2<T<T_1$, 
the two phenotypes with and 
without regulation, showed a large difference, 
contrasting the small difference 
in the RS1 phase.

In the RS2 phase, the two phenotypes were provided 
by using the first and second eigenmodes to express 
non-regulated and regulated phenotypes, respectively, 
where the switching path between the two phenotypes can be described 
by the first and second eigenmodes of the two endpoint phenotypes. 
A one-dimensional quarter-half shape switching 
path connected the two endpoint phenotypes in the 
two-dimensional space spanned by 
the first and second eigenvectors of the J2 genotype. 
This switching path was robust to 
perturbations, in the sense that any trajectories deviating 
from the path were attracted to the path. 
Evolutionary construction of this one-dimensional path met the requirements of plasticity against
regulatory changes and robustness in phenotypes.
Further, the low-dimensionality of the switching path
allowed for quick switching between two stable phenotypes
depending on the regulation.

To understand the evolutionary origin of the one-dimensional switching path, 
we applied a two-dimensional approximation to the free energy landscape 
for the evolved genotype in the RS2 phase. 
By only considering randomness 
in the free sites of two endpoint phenotypes, 
it was found that the free energy takes a minimum along a quarter-circle shape 
in two dimensions. 
The two endpoint phenotypes were 
located near the quarter-circle path, 
and the switching trajectories followed the valley of 
the free energies connecting the two endpoints.
In this case, the minima relate to the sites that were active and regulated. 
The cooperative evolution of the active, regulatory, and free sites provided stable expression of the endpoint 
phenotypes and robust switching paths.

Our findings suggest that low dimensionality plays a crucial role 
in achieving both stable expressions of two phenotypes and 
large conformational changes over a stable path. This 
leads to the acquisition of both robustness and plasticity. 
Constraints on adaptive changes in phenotypes upon environmental and 
evolutionary changes have recently gathered much attention 
\cite{Furusawa-KK_PRE, Sato2020, Sakata-Kaneko2020, Tang2021}. 
The constraint attracts a low-dimensional subspace 
within the high-dimensional space, supporting the robustness. 
Here, we demonstrated that the state change relevant to function 
is facilitated by the one-dimensionally constrained path on 
the two-dimensional plane, which allows large-amplitude plastic motion that is advantageous for functional changes.
Notably, this constrained path is already 
``prepared" as a relaxation path
during the evolution course (Fig.\ref{fig:gen_on_2dim}).

Previous studies have demonstrated that genotypes providing 
a single function by expressing a specific phenotype 
can evolve in the RS phase \cite{Sakata-Kaneko2020}. 
In our study, we found the transition that occurs in the RS phase 
for two-functional phenotypes. 
The genotypes that achieve switching between two functional 
phenotypes depending on the regulation were dominant 
in the RS2 phase, i.e., in the temperature region 
closer to the RSB within the RS phase.  
For the evolution to achieve more functions, 
further transitions within the RS phase can be 
expected. With such successive transitions, the genotype will approach the RSB 
transition point, where further plasticity will be achieved. This may be consistent with the 
observation of critical behavior in protein dynamics \cite{Tang}, 
wherein plasticity and 
robustness are compatible. 

Here, we did not impose any driving force to create the one-dimensional switching path;
rather, the evolution under the fitness defined by the two endpoint phenotypes 
resulted in genotypes that provide not only stable expression of the phenotypes 
but also robust and plastic switching. 
This observation presents the possibility of the evolutionary construction of 
proteins \cite{Togashi-Mikhailov} with allosteric effects
based on the binding ability of the active site, 
under conditions characterized by the RS phase,
in addition to synthetic approaches \cite{Raman2014}. 
Further analysis of interacting spin systems that achieve robust multiple functions is essential for the evolution of proteins and material design
\cite{Bornscheuer2001,Bahar2017}.

Investigation of the microscopic properties of 
evolved genotypes is an important future research direction. However, the focus of this study was on 
the extraction of macroscopic low-dimensional structures.
Frustration is a potential measure to characterize the genotype, 
which captures consistency in interactions, and the increase in frustration can 
indicate a rugged landscape. 
Generally, as the number of embedded patterns increases, 
the level of frustration in the interactions increases \cite{Hopfield1982, Amit1985}. 
We observed an increase in frustration in our model 
in comparison to the one-desirable phenotype case (see Supplement). 
In actual proteins, steric frustration can be utilized by multisubstrate enzymes 
to facilitate the rate-limiting product-release step \cite{Li2019}. 
Understanding the relationship between frustration and the number of embedded 
patterns may provide insights into the properties of real biomolecules. 

The evolutionary spin model considered in this study is rather simple and abstract. 
There is room to consider more realistic settings and 
discuss the generality of the results. 
For instance, several biological molecules have multiple regulatory or active sites, 
and their phenotype expression is more complicated. 
G protein-coupled receptors show dual ligand binding events where the binding of one ligand enhances that of the other \cite{Wootten2013, Lu2017}. Thiamine diphosphate in the two active sites of pyruvate dehydrogenase complex can communicate with each other over a distance of 20 angstroms using a proton to switch the conformation \cite{Frank2004}. The contribution of these kinds of cooperation to the evolution of robustness and plasticity needs to be revealed. 
In contrast to the global (all-to-all) coupling model, the study of models with spatially localized interactions is also important \cite{Hemery2015,Tlusty, Revoir}.

To conclude, we have shown that 
the stable expression and switching of phenotypes 
takes advantage of evolutionary constructed low-dimensional phenotypic constraints,
with which robustness and plasticity are compatible.
Our finding indicates that the 
evolution of low-dimensionality 
can be a unified view for the understanding of evolutional phenomena.

\appendix

\section{Derivation of the free energy density}

We introduce the equality
\begin{align}
\nonumber
1&=\int dm_k\delta\left(m_k-\frac{1}{\sqrt{N}}\sum_{i=1}^N\xi_i^kS_i\right)
\end{align}
for $k=1,2$ to the partition function with regulation as
\begin{align}
\nonumber
Z_+&=\int dm_1^+dm_2^+\sum_{\bm{S}|\bm{S}_{\cal R}^+\in\bm{S}_{\cal R}}\prod_{k=1}^2\delta\left(m_k-\frac{1}{\sqrt{N}}\sum_{i=1}^N\xi_i^kS_i\right)\exp\left[\sum_{k=1}^2
\frac{\beta\lambda_k}{2}\left(m_k^2-1\right)\right]\\
\nonumber
&=\int dm_1^+dm_2^+d\hat{m}_1^+d\hat{m}_2^+
\sum_{\bm{S}|\bm{S}_{\cal R}^+\in\bm{S}_{\cal R}}\prod_{k=1}^2\exp\left(-Nm_k\hat{m}_k+\sqrt{N}\hat{m}_k\sum_{i=1}^N\xi_i^kS_i+\frac{\beta\lambda_k}{2}(m_k^2-1)\right)\\
\nonumber
&=\int dm_1^+dm_2^+d\hat{m}_1^+d\hat{m}_2^+
\prod_{k=1}^2\exp\left(-Nm_k\hat{m}_k+\frac{\beta\lambda_k}{2}(m_k^2-1)\right)\\
&\hspace{1.0cm}\times\prod_{i=1}^{N}\left\{2\cosh(\sqrt{N}\sum_{k=1}^2\hat{m}_k\xi_i^k)\right\}
\times\left\{p_+^++p_-^+\right\}.
\end{align}
The integrals are implemented by utilizing the saddle point method.

For the without-regulation case,
we obtain
\begin{align}
\nonumber
Z_-&=\int dm_1^-dm_2^-d\hat{m}_1^-d\hat{m}_2^-
\sum_{\bm{S}|\bm{S}_{\cal R}\notin\bm{S}_{\cal R}^+}\prod_{k=1}^2\exp\left(-Nm_k\hat{m}_k+\sqrt{N}\hat{m}_k\sum_{i=1}^N\xi_i^kS_i+\frac{\beta\lambda_k}{2}(m_k^2-1)\right)\\
\nonumber
&=\int dm_1^-dm_2^-d\hat{m}_1^-d\hat{m}_2^-
\prod_{k=1}^2\exp\left(-Nm_k\hat{m}_k+\frac{\beta\lambda_k}{2}(m_k^2-1)\right)\\
&\hspace{1.0cm}\times\prod_{i=1}^{N}\left\{2\cosh(\sqrt{N}\sum_{k=1}^2\hat{m}_k\xi_i^k)\right\}
\times\left\{1-(p_+^-+p_-^-)\right\}.
\end{align}

\subsection{Free energy under assumptions {\bf A1}-{\bf A3}}

Under the assumptions A1 and A3,
$Z^+=Z^-$ holds, and we denote it as $Z_\pm$ given by
\begin{align}
\nonumber
Z_\pm&=\sum_{\bm{S}}\exp\left(\beta\overline{\lambda}\sum_{i<j}(\xi_i^1\xi_j^1+\xi_i^2\xi_j^2)S_iS_j\right)\\
\nonumber
&\sim\sum_{\bm{S}}\int dm_1dm_2
\delta\left(m_1-\frac{1}{\sqrt{N}}\sum_{i=1}^N\xi_i^1S_i\right)
\delta\left(m_2-\frac{1}{\sqrt{N}}\sum_{i=1}^N\xi_i^2S_i\right)
\exp\left(\displaystyle\frac{\beta\overline{\lambda}N}{2}
\left(m_1^2+m_2^2\right)\right)\\
\nonumber
&=\int dm_1dm_2d\hat{m}_1d\hat{m}_2
\exp(-N\hat{m}_1m_1-N\hat{m}_2m_2)\exp\left(\displaystyle\frac{\beta\overline{\lambda}N}{2}
\left(m_1^2+m_2^2\right)\right)\\
\nonumber
&\hspace{1.0cm}\times\prod_i2\cosh(\hat{m}_1\sqrt{N}\xi_i^1+\hat{m}_2\sqrt{N}\xi_i^2)\\
&=\int dm_1dm_2
\exp\left(-\displaystyle\frac{\beta\overline{\lambda}N}{2}
\left(m_1^2+m_2^2\right)+\sum_i\ln \cosh(\beta\overline{\lambda}\sqrt{N}(m_1\xi_i^1+m_2\xi_i^2))\right),
\end{align}
Here, we introduced the saddle point methods for the integral with respect to
$\hat{m}_1$ and $\hat{m}_2$.
Under the assumption {\bf A2} at sufficiently large system size,
the summation with respect to the components of the eigenvectors
can be replaced with the integral according to the Gaussian distribution.
as
\begin{align}
Z_\pm&=\int dm_1dm_2 \exp\left(-\displaystyle\frac{\beta\overline{\lambda}N}{2}
\left(m_1^2+m_2^2\right)+N\int Dz\ln\cosh\left(\beta\overline{\lambda}\sqrt{m_1^2+m_2^2}z\right)\right).
\end{align}
By introducing the saddle point method to the integrals of 
$m_1$ and $m_2$,
we obtaine the approximated free energy $f_{\mathrm{app}}$.

\begin{acknowledgments}

The authors thank Qian Yuan Tang and Tuan Pham
for helpful comments and discussions.
This work is partially supported by Grant in-Aid for Scientific Research (A) (20H00123) from the Ministry of Education, Culture, Sports, Science, and Technology (MEXT) of Japan.
KK is also supported by the Novo Nordisk Foundation.

\end{acknowledgments}

\newpage
%

\end{document}